\documentclass[11pt,english]{article}
\pdfoutput=1
\usepackage[T1]{fontenc}
\usepackage{amssymb}
\usepackage{amsmath}
\usepackage{cite}
\usepackage{appendix}
\usepackage[unicode=true,
 bookmarks=true,bookmarksnumbered=false,bookmarksopen=false,
 breaklinks=false,pdfborder={0 0 1},backref=false,colorlinks=true]
 {hyperref}
\usepackage{breakurl}
\usepackage[hang,flushmargin]{footmisc}
\usepackage{color}
\usepackage{fullpage}
\usepackage{graphicx}
\usepackage{bm}

\graphicspath{{draft-figures/}}

\usepackage{chngcntr}
\counterwithin{equation}{section}

\hypersetup{pdftitle={Covariant Predictions for Planck-Scale Features in Primordial Power Spectra},
 pdfauthor={Aidan Chatwin-Davies, Achim Kempf, Petar Simidzija},
 citecolor=black,linkcolor=black,urlcolor=black}
 
\newcommand{\email}[1]{\href{mailto:#1}{\nolinkurl{#1}}}

\makeatletter
\def\simgt{\mathrel{\lower2.5pt\vbox{\lineskip=0pt\baselineskip=0pt
           \hbox{$>$}\hbox{$\sim$}}}}
\def\simlt{\mathrel{\lower2.5pt\vbox{\lineskip=0pt\baselineskip=0pt
           \hbox{$<$}\hbox{$\sim$}}}}
\makeatother

\renewcommand{\Ref}[1]{Ref.~\cite{#1}}
\newcommand{\Fig}[1]{Fig.~\ref{#1}}
\newcommand{\Eq}[1]{Eq.~\eqref{#1}}
\newcommand{\Sec}[1]{Sec.~\ref{#1}}
\newcommand{\App}[1]{App.~\ref{#1}}

\DeclareMathOperator{\sech}{sech}
\DeclareMathOperator{\csch}{csch}

\newcommand{\bra}[1]{\langle #1|}
\newcommand{\ket}[1]{|#1\rangle}

\newcommand{\ketbra}[2]{| #1 \rangle \langle #2 |}
\newcommand{\ip}[2]{\langle #1 , #2 \rangle}

\newcommand{\mrm}[1]{\mathrm{#1}}
\newcommand{\dee}{\mathrm{d}}

\renewcommand{\Re}{\mathrm{Re}}
\renewcommand{\Im}{\mathrm{Im}}

\newcommand{\calR}{\mathcal R}
\newcommand{\calT}{\mathcal T}

\newcommand{\Mpl}{M_\mrm{Pl}}

\usepackage{xcolor}

\definecolor{purple}{rgb}{0.5,0,0.5}

\makeatletter

\def\lambdabar{\protect\@lambdabar}
\def\@lambdabar{%
\relax
\bgroup
\def\@tempa{\hbox{\raise.73\ht0
\hbox to0pt{\kern.25\wd0\vrule width.5\wd0
height.1pt depth.1pt\hss}\box0}}%
\mathchoice{\setbox0\hbox{$\displaystyle\lambda$}\@tempa}%
{\setbox0\hbox{$\textstyle\lambda$}\@tempa}%
{\setbox0\hbox{$\scriptstyle\lambda$}\@tempa}%
{\setbox0\hbox{$\scriptscriptstyle\lambda$}\@tempa}%
\egroup
}

\makeatother

\begin{document}

\interfootnotelinepenalty=10000
\baselineskip=18pt

\hfill

\vspace{2cm}
\thispagestyle{empty}
\begin{center}
{\LARGE \bf
Covariant Predictions for Planck-Scale Features in\\Primordial Power Spectra}\\
\bigskip\vspace{1cm}{
{\large Aidan Chatwin-Davies${}^{a,b}$, Achim Kempf$\,{}^{c}$, and Petar Simidzija${}^{a}$}
} \\[7mm]
 {\it ${}^a$Department of Physics and Astronomy, University of British Columbia\\[-1mm]
 6224 Agricultural Road, Vancouver, BC, V6T 1Z1, Canada\\[1.5mm]
 ${}^b$Institute for Theoretical Physics, KU Leuven\\[-1mm]
 Celestijnenlaan 200D B-3001 Leuven, Belgium \\[1.5 mm]
 ${}^c$Department of Applied Mathematics, University of Waterloo\\[-1mm]
 Waterloo, ON, N2L 3G1, Canada}
 \let\thefootnote\relax\footnote{\noindent e-mail: \email{achatwin@phas.ubc.ca}, \email{akempf@pitp.ca}, \email{psimidzija@phas.ubc.ca}} \\
 \bigskip\vspace{0.5cm}{\today}
 \end{center}
\bigskip
\centerline{\large\bf Abstract}
\begin{quote} \small
In this companion to our letter (arXiv:2208.10514), we elaborate the full details of the predicted corrections to the primordial scalar and tensor power spectra that arise from quantum gravity-motivated, natural, covariant ultraviolet cutoffs. We implement these cutoffs by covariantly restricting the fields which are summed over in the path integrals for the primordial correlators, and we discuss in detail the functional analytic techniques necessary for evaluating such path integrals. Our prediction, which is given in terms of measured cosmological parameters and without assuming any particular inflationary potential, is that the corrections take the form of small oscillations which are superimposed on the conventional power spectra. The frequency of these oscillations only depends on the location of the cutoff scale and the first inflationary slow-roll parameter, while the amplitude and phase are also moderately sensitive to how smoothly the cutoff turns on. The specificity of the new predictions offers an opportunity to significantly enhance experimental sensitivity in observations of the cosmic microwave background and large-scale structure. This may be used to place ever higher bounds on the scale at which quantum gravity effects become important in quantum field theory or may even provide positive evidence for quantum gravity effects.
\end{quote}

\setcounter{footnote}{0}

\newpage
\tableofcontents
\newpage

\section{Introduction}

The development of quantum gravity has been impeded by the lack of experimental access to the Planck scale. For example, the peak energy of the Large Hadron Collider is still about 15 orders of magnitude below the Planck energy, and so the quantum gravitational regime remains well out of the reach of accelerator experiments. 
The numbers are more favorable, however, in cosmology.
This is because, according to the standard model of cosmology, the inhomogeneities in the Cosmic Microwave Background (CMB) originated in quantum fluctuations of modes which froze when they exceeded the Hubble length during inflation. Since the Hubble length at that time was only about 5 to 6 orders of magnitude larger than the Planck length, Planck-scale effects in the CMB should be correspondingly less suppressed and could perhaps even become observable.

The question arises, therefore, as to what exact signature of potential Planck-scale effects in the CMB to predict, so that experimental efforts can be guided towards probing these predictions \cite{Chluba:2015bqa,Slosar:2019gvt}. The candidate theories for quantum gravity differ strongly in their description of physics at the Planck scale and therefore, in principle, each theory could yield its own predictions \cite{Rovelli:1997qj,WittenStrings,Carlip:2015asa,Loll:2022ibq}. However, just below the Planck energy scale, each candidate theory for quantum gravity must quite quickly reduce to quantum field theory on curved spacetime in order to be consistent with the current standard model of cosmology. This indicates that, at the Hubble scale during inflation, some 5 or 6 orders of magnitude from the Planck scale, only the most dominant features of Planck-scale physics should have been able to leave an imprint in the quantum fluctuations of the inflaton and metric modes that froze at that time. 

This leads to the question as to what is the most dominant impact of Planck-scale physics on quantum field theory in curved spacetime at those scales where quantum field theory is still a good enough model to describe inflation. Since candidate theories of quantum gravity tend to predict the presence of some form of natural ultraviolet (UV) cutoff \cite{Garay:1994en,Hossenfelder:2012jw}, it is natural to conjecture that
quantum field theory close to the Planck scale is modified predominantly by the presence of an ultraviolet cutoff, which may be hard or soft. The challenge is then to predict the impact that such a cutoff in quantum field theory in curved spacetime would have on the predictions for the CMB. 
The literature on this question has been mostly working with noncovariant ultraviolet cutoffs and associated modified dispersion relations, see, e.g., \cite{Padmanabhan:1988jp,Padmanabhan:1988se,Jacobson:1999zk,Kempf:2000ac,Martin:2000xs,Brandenberger:2000wr,Niemeyer:2000eh,Brandenberger:2001zqv,Easther:2001fi,Kempf:2001fa,Easther:2001fz,Brandenberger:2002hs,Easther:2002xe,Danielsson:2002kx,Brandenberger:2004kx,Sriramkumar:2004pj,Greene:2004np,Shiu_2005,Easther:2005yr,Tawfik:2015rva,Ali:2015ola,Skara:2019uzz,Frob:2012ui,Frob:2014cza,Kaloper:2002uj}.
For cutoff-free models, see, e.g., \cite{Calcagni:2016ofu,Calcagni:2017via,Modesto:2022asj,Calcagni:2022tuz}.
There has also been significant interest in and progress made on this front and related questions within the field of Loop Quantum Cosmology; see, e.g., \cite{deBlas:2016puz,Ashtekar:2020gec,Agullo:2021oqk,Ashtekar:2021izi,Navascues:2021mxq,Navascues:2021qcp}.

In previous work, we studied the case of a hard natural ultraviolet cutoff that is covariant \cite{Kempf:2012sg,Chatwin-Davies:2016byj}.
That the cutoff is covariant is important to ensure that the predictions arise only from the presence of the UV cutoff and are uncontaminated by the breaking of symmetries.
The cutoff is enacted on the spectrum of the scalar field's spacetime d'Alembertian, and it has an innate information theoretic interpretation as a cutoff on the field's density of degrees of freedom in spacetime.
We then presented a proof-of-principle calculation to illustrate how the apparatus could be used to compute the signature that a covariant UV cutoff would leave in the spectrum of inflationary perturbations in the CMB \cite{Chatwin-Davies:2016byj}.

In the present paper, we build on this ansatz and explicitly calculate the correction that a natural covariant UV cutoff at (or near) the Planck scale produces for inflationary primordial power spectra, assuming slow-roll parameters that arise from observation.
We focus on the scalar power spectrum, i.e., the dimensionless power spectrum of the comoving curvature perturbation, $\Delta_\mathcal{R}^2$, since it has already been characterized with ample amounts of observational data \cite{Planck:2013oqw,Planck:2015mrs,Planck:2018nkj}, with more data from the CMB and large scale structure surveys on the way \cite{Slosar:2019gvt}.
A natural covariant UV cutoff also produces a correction to the as-yet unobserved primordial spectrum of tensor perturbations, $\Delta_\mathcal{T}^2$, which we also compute.

We assume that inflation is driven by a single inflaton field, but beyond this assumption our calculation is model-independent, in the sense that we make no assumptions about the inflaton's potential.
The only input required for the calculation is the background Friedmann-Lema{\^i}tre-Robertson-Walker (FLRW) Hubble parameter which describes the inflationary phase, which we fix using measured values of slow-roll parameters and the parameters that describe $\Delta_\calR^2$ (see \Eq{eq:Hubble-eff}).
We choose the Bunch-Davies vacuum state for the cosmological perturbations so as to minimally diverge from the standard inflationary calculation of $\Delta_\calR^2$.
It is possible to accommodate a different choice of vacuum state within our framework; however, the resulting predictions may change significantly as a consequence.

We find that a natural covariant UV cutoff produces small oscillations which are a function of the comoving momentum $k$, superimposed on top of the uncorrected power spectrum, as illustrated in \Fig{fig:prediction}a.
The predicted effect depends only on two parameters: 1) the first slow-roll parameter $\epsilon_\star\equiv \epsilon(k_\star)$ at some fiducial momentum scale $k_\star$, and 2) the dimensionless ratio $\sigma(k)$ of the cutoff length, $\ell_c$, to the Hubble length, which varies with the comoving wave number $k$ during the slow roll.
In terms of observational parameters, $\sigma(k)$ is given by
\begin{align}
    \sigma(k) \equiv \left. \frac{H}{\Omega} \right|_{aH=k} =  \frac{\Mpl}{\Omega}\sqrt{\pi A_s\epsilon_\star}\left(\frac{k}{k_\star}\right)^{-\epsilon_\star}.
\end{align}
where $\Omega \equiv 1/\ell_c$ is the energy scale associated with the cutoff and $A_s$ is the scalar perturbation amplitude.
Our finding is that, for a sharp cutoff, the relative change in the power spectrum as a function of the comoving mode $k$ is given by
\begin{equation}
   \frac{\delta \Delta^2_\calR}{\Delta^2_\calR}=
    \mathcal{C}
    \frac{\sigma(k)^{3/2}}{\ln(\sigma(k)/2)} \sin\left(\omega(k)\,  \sigma(k)\right),
\end{equation} 
where $\mathcal{C}=0.8796...$ is a numerical constant and where we have defined
\begin{align}
    \omega(k) &\equiv \frac{1}{\sigma(k)^2}\ln\frac{\sigma(k)e}{2}.
\end{align}
While the oscillations' amplitude and phase depend mildly on fine-grained details of the cutoff, such as how gradually it turns on, the oscillations' frequency is a robust prediction which is essentially independent of the hardness or softness of the cutoff.
The only free parameters in this prediction are 
the first slow-roll parameter, $\epsilon_\star$ (which is constrained by CMB observations \cite{Planck:2018jri} to $0 < \epsilon_\star\lesssim 0.004$), and the precise energy scale of the UV cutoff, $\Omega$.

For a fixed value of $\epsilon_\star$, the smaller the value of $\Omega$, the smaller is the peak oscillation frequency and the larger is the amplitude of the oscillations, resulting in a larger imprint on the primordial power spectrum; see Fig. \ref{fig:Amp_and_Nosc} for illustration.
Consequently, existing and future measurements should be able to place bounds on $\Omega$, the scale at which quantum gravity effects become important in inflation, in relation to $\epsilon_\star$.
In particular, an eventual measurement of $\epsilon_\star$, such as through the amplitude of the CMB tensor modes, would strenghthen the bound on $\Omega$.
Furthermore, an explicit measurement of the predicted oscillations would fix both the values of $\epsilon_\star$ and $\Omega$.

Since our calculations involve a variety of techniques, we have opted to be liberal when covering background material and in the expositions.
In \Sec{sec:cosmological-perturbations}, we therefore begin by briefly reviewing relevant aspects of the theory of cosmological perturbations and the computation of primordial power spectra while establishing notation and definitions in preparation for our subsequent calculations.
Next, in \Sec{sec:covariant-cutoff}, we review the definition and implementation of the covariant natural UV cutoff which we consider.
We demonstrate how such a cutoff produces a correction to the power spectrum of a scalar field in de Sitter spacetime and we explain how to extend the result to slowly rolling FLRW spacetimes.
In \Sec{sec:prediction}, we take the scalar field to be the comoving curvature perturbation, and we compute the correction to its primordial power spectrum assuming single field inflation and realistic slow-roll parameters.
We also discuss the correction to the tensor spectrum here.
Finally, we end with a summary and discussion in \Sec{sec:discussion}.
The most intensive of calculational details appear in the subsequent appendices.

\section{Cosmological perturbations}
\label{sec:cosmological-perturbations}

The most remarkable success of the theory of inflation is its ability to predict a primordial power spectrum which is in quantitative agreement with observed large scale fluctuations in the universe, as seen, for example, in the cosmic microwave background. Let us briefly review how this primordial power spectrum is computed.

In the simplest model of inflation, which we consider here, the spacetime metric $g_{\mu\nu}$ is coupled to a scalar field $\phi$, called the inflaton, via the action
\begin{align}\label{eq:action}
    S = \frac{\Mpl^2}{16\pi} \int \dee^4 x \sqrt{-g} R 
    - \int \dee^4 x \sqrt{-g} \left[\frac{1}{2}\partial_\mu \phi\partial^\mu \phi +V(\phi)\right],
\end{align}
where $\Mpl \equiv 1/\sqrt{G}$ is the Planck mass, and we work in units $c=\hbar=1$. On the largest scales, the universe is nearly spatially homogeneous and isotropic, and hence the metric and inflaton field can be written as
\begin{align}
    g_{\mu\nu}(\eta,\bm x) &= a^2(\eta)\eta_{\mu\nu}+ h_{\mu\nu}(\eta, \bm x),\label{eq:metric}\\
    \phi(\eta, \bm x) &= \bar\phi(\eta)+ \delta \phi (\eta,\bm x)\label{eq:inflaton}.
\end{align}
The first terms describe a spatially flat FLRW cosmology with a scale factor $a(\eta)$ and a spatially constant background field $\bar\phi(\eta)$, which both depend only on the conformal time $\eta$, while the second terms allow for deviations from spatial homogeneity and isotropy. We will assume that these deviations are small and we will treat them quantum mechanically, while the dominant background pieces will be treated classically.

\subsubsection*{Background}

Substituting \eqref{eq:metric} and \eqref{eq:inflaton} into the action \eqref{eq:action}, and keeping only leading order terms, one obtains the equations of motion for the background fields
\begin{align}
    H^2 &= \frac{8\pi}{3\Mpl^2}\left(\frac{1}{2}\bar\phi'^2 + V(\bar\phi)\right),\label{eq:background_Friedmann}\\
    \dot H &= -\frac{4\pi\bar\phi'^2}{\Mpl^2a^2},\label{eq:second_Friedmann}\\
    \bar\phi'' &= -2a H\bar\phi'-a^2 V_\phi(\bar\phi),\label{eq:background_scalar}
\end{align}
where $V_\phi\equiv dV/d\phi$, $H\equiv \dot a/a = a'/a^2$ is the Hubble parameter, primes denote derivatives with respect to conformal time, $\eta$, and dots are derivatives with respect to cosmic time, $t$. 

The theory of inflation postulates that in the very early universe, perhaps by some abnormally large quantum fluctuation, the inflaton field found itself at a value where the potential is large (close to the Planck scale) but the gradient of the potential is given by a much lower scale. The equations of motion \eqref{eq:background_Friedmann} and \eqref{eq:background_scalar} give the resulting background dynamics: $\bar\phi(\eta)$ slowly rolls down the potential from its large value, while $a(\eta)$ experiences a period of highly accelerated expansion, characterized by a Hubble parameter $H(\eta)$ which is slowly decreasing in time. More precisely, one can quantify the rate of change of $H$ via the \textit{slow-roll parameters}
\begin{align} \label{eq:slow_roll_params}
    \epsilon \equiv -\frac{\dot H}{H^2}, \quad \delta \equiv \frac{\ddot H}{2H\dot H}.
\end{align}
Slow-roll inflation is characterized by the conditions $\epsilon \ll 1 $ and $\delta \ll 1$. 

\subsubsection*{Perturbations}

Now let us consider the fluctuations $\delta\phi$ and $h_{\mu\nu}$ on top of this spatially homogeneous and isotropic classical background. To obtain the dynamics of these fields, one again substitutes \eqref{eq:metric} and \eqref{eq:inflaton} into the action \eqref{eq:action}, this time keeping terms up to second order in the perturbations; hence, the perturbations are free fields. However, even in the absence of interactions, there is a challenge in quantizing these fields due to the fact that the theory is diffeomorphism-invariant, i.e. there is gauge freedom associated with the choice of coordinates. The quantization of gauge theories is complicated by the fact that for a gauge invariant action such as \eqref{eq:action}, it is not manifest which fields constitute the physical degrees of freedom of the theory and which fields can simply be gauged away. A careful analysis shows that after gauge fixing, the only remaining degrees of freedom in the perturbations are a scalar, $\calR(x)$---the Mukhanov-Sasaki variable---and two degrees of freedom associated with a transverse, traceless, symmetric tensor, $h_{ij}(x)$. The gauge-fixed action is given by $S = S_\calR + S_h$, where \cite{Mukhanov:1990me}
\begin{align}
    S_\calR &= -\frac{1}{2}\int \dee^4 x \, z^2 \eta^{\mu\nu} \partial_\mu \calR\partial_\nu \calR,\\
    S_h &= -\frac{\Mpl^2}{64\pi} \int \dee^4 x\, a^2\eta^{\mu\nu}\partial_\mu h_{ij}\partial_\nu h_{ij}.\label{eq:S_h}
\end{align}
If we ignore the tensor structure, we see that the field $h_{ij}$---which describes primordial gravitational waves with two polarizations---simply has the dynamics of a massless scalar field in a FLRW spacetime with scale factor $a(\eta)$. 
On the other hand, the Mukhanov-Sasaki variable experiences a modified scale factor, $z(\eta)$, defined in terms of the background fields as
\begin{align} \label{eq:z}
    z \equiv \frac{a\dot{\bar\phi}}{H} = \frac{a^2\dot{\bar\phi}}{a'} =\frac{a\bar\phi'}{a'}.
\end{align}
Notice that the second Friedmann equation \eqref{eq:second_Friedmann} implies $z = (\Mpl^2\epsilon/4\pi)^{1/2}a$. Hence, if the first slow-roll parameter $\epsilon$ is constant, $z$ is simply proportional to $a$, and thus the Hubble parameter $\dot z/z$ associated with the modified scale factor is equal to the Hubble parameter $H=\dot a/a$ associated with the ``true'' scale factor. More generally, $\epsilon$ is not constant, but rather varies as \cite{weinberg2008cosmology}
\begin{align}
    \dot \epsilon = 2H\epsilon(\epsilon+\delta),
\end{align}
which implies
\begin{align} \label{eq:z-Hubble}
    \frac{\dot z}{z} = H(1+\epsilon+\delta).
\end{align}

\subsubsection*{Canonical quantization}

Let us now canonically quantize the scalar and tensor perturbations by expanding them in terms of spatial Fourier modes. We obtain
\begin{align}
    \calR(x) & = \frac{1}{z(\eta)}\int \frac{\dee^3 \bm k}{(2\pi)^{3/2}}e^{i\bm k\cdot\bm x}u_{\bm k}(\eta)a_{\bm k}^\dagger+h.c.
    \\
    h_{ij}(x) & = \frac{\sqrt{16\pi}}{a(\eta)\Mpl}\sum_{\lambda=1}^2\int \frac{\dee^3 \bm k}{(2\pi)^{3/2}}e^{i\bm k\cdot\bm x}\epsilon_{ij}(\hat{\bm k},\lambda)v_{\bm k}(\eta)b_{\bm k,\lambda}^\dagger+h.c.
\end{align}
Here, $a^\dagger, b^\dagger$, and their adjoints are canonically commuting creation and annihilation operators, and $\epsilon_{ij}(\hat{\bm k},1)$ and $\epsilon_{ij}(\hat{\bm k},2)$ are two linearly independent, symmetric, traceless, and transverse polarization tensors. The mode functions $u_{\bm k}(\eta)$ and $v_{\bm k}(\eta)$ are harmonic oscillators with time-dependent frequencies
\begin{align}
    u_{\bm k}''+\left(k^2 - \frac{z''}{z}\right)u_{\bm k}&=0, \label{eq:mode-fcn-eom-scalar}\\
    v_{\bm k}''+\left(k^2 - \frac{a''}{a}\right)v_{\bm k}&=0. \label{eq:mode-fcn-eom-tensor}
\end{align}
In the slow-roll regime, both $a(\eta)$ and $z(\eta)$ approach $1/(-\eta)$ for large negative values of $\eta$, and hence in this limit both equations simply reduce to that of a harmonic oscillator with frequency $k = |\bm k|$. A natural choice of vacuum state is the Bunch-Davies vacuum, which corresponds to setting the usual positive frequency initial conditions
\begin{align}
    u_{\bm k}(\eta) 
    \rightarrow 
    \frac{1}{\sqrt{2k}}e^{-i \omega \eta},
    \quad\text{and}\quad
    v_{\bm k}(\eta) 
    \rightarrow
    \frac{1}{\sqrt{2k}}e^{-i \omega \eta},
    \label{eq:u_v_initial_condition}
\end{align}
at $\eta\rightarrow -\infty$. We denote this vacuum by $\ket 0$ and assume that the field starts out in this state. 

The quantum fluctuations of the scalar and tensor perturbations can be quantified in terms of their two-point correlators. The equal time correlators are
\begin{align}
    G_\calR(\eta, \bm x) &\equiv \bra 0 \calR(\eta, \bm x)\calR(\eta, 0)\ket 0 
    = 
    \frac{1}{z(\eta)^2}
    \int \frac{\dee^3 \bm k}{(2\pi)^3} e^{-i\bm k \cdot \bm x} |u_{\bm k}(\eta)|^2,
    \\
    G_h^{ij,kl}(\eta, \bm x) &\equiv
    \bra 0 h_{ij}(\eta, \bm x)h_{kl}(\eta, 0)\ket 0 
    = 
    \frac{16\pi}{a(\eta)^2\Mpl^2}
    \int \frac{\dee^3 \bm k}{(2\pi)^3} e^{-i\bm k \cdot \bm x}
    |v_{\bm k}(\eta)|^2
    \Pi_{ij,kl}(\hat{\bm k}),
\end{align}
where we define the quantity $\Pi_{ij,kl}(\hat{\bm k}) \equiv \sum_\lambda \epsilon_{ij}(\hat{\bm k},\lambda)\epsilon_{kl}^*(\hat{\bm k},\lambda)$. The Fourier transforms of the two-point correlators are
\begin{align}
    G_\calR(\eta,\bm k)
    &\equiv \frac{1}{(2\pi)^3}\int \dee^3\bm x e^{i\bm k\cdot\bm x}G_\calR(\eta, \bm x) 
    = 
    \frac{|u_{\bm k}(\eta)|^2}{(2\pi)^3z(\eta)^2},\label{eq:Greens_function_R}
    \\
    G_h^{ij,kl}(\eta,\bm k) 
    &\equiv \frac{1}{(2\pi)^3}\int \dee^3\bm x e^{i\bm k\cdot\bm x}G_h^{ij,kl}(\eta, \bm x)
    =
    \frac{16\pi|v_{\bm k}(\eta)|^2}{(2\pi)^3a(\eta)^2\Mpl^2}\Pi_{ij,kl}(\hat{\bm k}). \label{eq:Greens_function_h_ij}
\end{align}
Notice that the tensor structure of the quantity $G_h^{ij,kl}$ is purely kinematic and that all of the dynamics is contained in the single function $v_{\bm k}(\eta)$. This is simply the statement that the two linearly independent tensor helicities have the same vacuum fluctuation amplitudes and each fluctuate as a free scalar field.
As shorthand, let $G_h$ denote $G_h^{ij,kl}$ modulo its kinematic tensor structure, i.e.
\begin{equation} \label{eq:Greens_function_h}
    G_h(\eta,\bm k) \equiv \frac{16\pi|v_{\bm k}(\eta)|^2}{(2\pi)^3a(\eta)^2\Mpl^2}.
\end{equation}

\subsubsection*{Primordial power spectra}

The scalar and tensor primordial power spectra are respectively defined as
\begin{align}
    \Delta_\calR^2(k)&\equiv \frac{k^3}{2\pi^2z(\eta_k)^2}|u_{\bm k}(\eta_k)|^2, \label{eq:scalar-fluc-spec}
    \\
    \Delta_\calT^2(k)&\equiv \frac{k^3}{2\pi^2a(\eta_k)^2}\frac{64\pi}{\Mpl^2}|v_{\bm k}(\eta_k)|^2,
\end{align}
where $\eta_k$ is the comoving time at which modes with comoving momentum of magnitude $k = |{\bm k}|$ cross the Hubble horizon, i.e. it is defined as the solution to the equation
\begin{align}
    k = a(\eta_k)H(\eta_k).
\end{align}
We will make use of a few alternate expressions for the primordial power spectra. In a slowly rolling spacetime, it is possible to obtain approximations for the mode functions $u_{\bm k}$ and $v_{\bm k}$ (or equivalently the two-point functions) at horizon crossing, i.e. at $\eta=\eta_k$, in terms of the value of the Hubble parameter. This results in the following expressions for the power spectra (see e.g. \cite{dodelson2003modern}):
\begin{align}
    \Delta_{\calR}^2(k) &= \frac{H^2}{\pi\epsilon \Mpl^2}\Big|_{\eta=\eta_k}, \label{eq:scalar-spectrum} \\
    \Delta_{\calT}^2(k) &= \frac{16H^2}{\pi \Mpl^2}\Big|_{\eta=\eta_k}. \label{eq:tensor-spectrum}
\end{align}
Notice that since the Hubble parameter and the slow-roll parameter $\epsilon$ vary slowly during inflation, we expect $\Delta_\calR$ and $\Delta_\calT$ to vary only mildly with $k$---the primordial power spectra should be roughly \textit{scale invariant}. Indeed, observations of cosmic perturbations indicate the scalar spectrum to be of the form
\begin{align} \label{eq:scalar-spectrum-pheno}
    \Delta_{\mathcal R}^2(k) = A_s\left(\frac{k}{k_\star}\right)^{n_s-1},
\end{align}
where, at the pivot scale $k_\star = 0.05~\mrm{Mpc}^{-1}$, the amplitude $A_s$ and spectral tilt $n_s$ are observed at the values \cite{Planck:2018vyg}
\begin{align}
    A_s &= (2.10\pm0.03)\times 10^{-9}, \quad\\
    n_s &= 0.966 \pm 0.004.
\end{align}
Similarly, the tensor spectrum can be empirically fit to the curve
\begin{align}
\Delta_\calT^2(k)=A_t\left(\frac{k}{k_\star}\right)^{n_t}.
\end{align}
Since tensor fluctuations have not yet been observed, the values of $A_t$ and $n_t$ are unknown, but the tensor-to-scalar ratio $r\equiv A_t/A_s$ is constrained to $r<0.06$ \cite{Planck:2018jri}.

Finally, note that using equations \eqref{eq:Greens_function_R} and \eqref{eq:Greens_function_h}, we can write the power spectra $\Delta_\calR$ and $\Delta_\calT$ in terms of two-point correlators as
\begin{align}
    \Delta_{\calR}^2(k) &= 4\pi k^3 G_\calR(\eta_k,k), \label{eq:scalar-spec-from-2pt}\\
    \Delta_{\calT}^2(k) &= 8\pi k^3 G_h(\eta_k,k).
\end{align}
(The extra factor of 2 in the tensor power spectrum comes from there being two linearly independent gravitational wave polarizations \cite{dodelson2003modern}.)
The advantage of this rewriting is that the two-point functions can be expressed in a manifestly covariant manner using a path integral. The goal of this paper will be to impose a covariant high energy cutoff on this path integral and to study this as a simple model of the way in which Planck-scale physics might affect inflationary power spectra.

\section{Covariant ultraviolet cutoffs}
\label{sec:covariant-cutoff}

In this section, we explain the machinery that we use to calculate predictions for Planckian corrections to CMB power spectra.
We begin by reviewing the class of covariant natural UV cutoffs that we work with.
We then give a detailed description of how to impose such a cutoff on the fluctuation spectrum of a quantized scalar field in de Sitter spacetime.
Finally, we use the de Sitter result to obtain the cutoff fluctuation spectrum for near-de Sitter FLRW spacetimes. 

The kinematics of covariant natural ultraviolet cutoffs are discussed in detail in Refs.\cite{Kempf:1999xt,Kempf:2003qu,Kempf:2009us,Kempf:2010rx,Kempf:2012sg}.
An outline of the de Sitter calculation was given in \Ref{Chatwin-Davies:2016byj}, and here we elaborate the calculation in full while making important refinements along the way.

\subsubsection*{Definition and implementation}

Starting with a quantum field theory on a curved spacetime, our aim is to model gravitational corrections to the effective theory via a natural UV cutoff, and we wish to do so covariantly.
Conceptually, our approach is to suppress field configurations in the quantum field theoretic path integral that lie beyond the Planck scale, which we parameterize in terms of eigenfunctions and eigenvalues of the field's d'Alembertian.

Concretely, let $(\mathcal{M},g)$ be a Lorentzian manifold $\mathcal{M}$ with metric $g$ that supports a real scalar quantum field $\hat \phi$.
Let $\Box$ denote the field's d'Alembertian,
\begin{equation}
    \Box = \frac{1}{\sqrt{-g}}\partial_\mu \left( \sqrt{-g} g^{\mu \nu} \partial_\nu \,\cdot \right),
\end{equation}
and let us suppose that any boundary conditions have been appropriately chosen so that $\Box$ is self-adjoint.
Given a non-negative function $f$, we define a cutoff on a field configuration $\phi(x)$ via a linear combination of projectors onto the eigenspaces of $\Box$:
\begin{equation} \label{eq:cutoff}
    f(\Box) ~ : ~ \phi(x) ~ \mapsto ~ \sum_{\lambda \in \mrm{spec}\,\Box} f(\lambda) \langle \psi_\lambda, \, \phi \rangle \psi_\lambda(x)
\end{equation}
Here, $\psi_\lambda$ denotes the eigenfunction of $\Box$ with eigenvalue $\lambda$, and $\langle \, \cdot \; , \, \cdot \, \rangle$ is the $L^2(\mathcal{M})$ inner product.
For example, if
\begin{equation} \label{eq:sharp-f}
    f(\lambda) = \theta(\Omega^2 - |\lambda|),
\end{equation}    
where $\theta$ is the Heaviside step function, then $f(\Box)$ is a sharp cutoff that projects fields onto the subspace spanned by eigenfunctions whose eigenvalues' magnitudes are less than $\Omega^2$.
Cutoffs of the form \eqref{eq:cutoff} are manifestly covariant; they do not depend on a choice of coordinates for $\mathcal{M}$, as they are specified entirely in terms of the spectrum of $\Box$, which is itself just a set of real numbers that depends on $\mathcal{M}$ alone.

Flat spacetime is an illustrative example.
Choosing $\mathcal{M} = \mathbb{R}^{1,d}$ and usual Cartesian coordinates $(t,{\bm x})$, the d'Alembertian is $\Box = -\partial_t^2 + \partial_j \partial^j$ and its eigenfunctions are plane waves, $\psi_{k}(x) = e^{-i k^0 t + i {\bm k}\cdot{\bm x}}$.
The corresponding eigenvalues are $(k^0)^2 - {\bm k}^2$, and so a sharp cutoff like \eqref{eq:sharp-f} removes Fourier contributions to a field configuration from plane waves whose spacetime momentum-squared is greater than $\Omega^2$ or less than $-\Omega^2$.
For a given field configuration $\phi(x)$ with Fourier transform $\tilde \phi(k)$, the action of
\begin{equation}
P_\Omega \equiv \theta(\Omega^2 - |\Box|)
\end{equation}
is therefore
\begin{equation} \label{eq:bandlimited-field-flat}
P_\Omega \phi(x) = \int_{|k_\mu k^\mu| \leq \Omega^2} \frac{\dee^{d+1}k}{(2\pi)^{d+1}} ~ \tilde \phi(k)~  e^{i k_\nu x^\nu}.
\end{equation}

The quantity $\Omega$ plays the role of a short-distance cutoff.
Moreover, $\Omega$ admits a natural information theoretic interpretation as a covariant bandlimit on the density of degrees of freedom in spacetime, in the sense of Nyquist-Shannon sampling theory \cite{shannon1998mathematical,NyquistReprint}.
Given a Riemannian manifold, one can view a \emph{conventional} bandlimit as a cutoff on the spectrum of the Laplacian, $\bigtriangleup$.
For example, for functions on $\mathbb{R}$, with $\bigtriangleup = -\partial_x^2$ and eigenfunctions $e^{ikx}$, a cutoff $\Lambda$ restricts $k^2 \leq \Lambda^2$. 
\emph{Bandlimited functions} are then functions whose Fourier transforms are compactly supported in a finite interval $[-\Lambda, \Lambda]$, and the maximum Fourier frequency $\Lambda$ is the \emph{bandlimit}, or equivalently here, the functions' \emph{bandwidth}.
A sampling theorem then applies: a bandlimited function can be perfectly reconstructed everywhere on the real line knowing only its values at a discrete set of sample points whose average density is greater than or equal to $\Lambda/\pi$ \cite{LandauSampling}.
With appropriate modifications, versions of the sampling theorem above generalize to $\mathbb{R}^d$ and to Riemannian manifolds \cite{LandauSampling,PesensonSampling}.

The covariant cutoff $\Omega$ on $\mathbb{R}^{1,d}$ is somewhat different, mainly because the set of allowed eigenvalues $|(k^0)^2 - {\bm k}^2| \leq \Omega^2$ is not compact.
Nevertheless, each partial Fourier transform of a covariantly bandlimited field such as \eqref{eq:bandlimited-field-flat} enjoys a sampling theorem.
That is, consider taking a partial Fourier transform of $P_\Omega \phi$ with respect to $\bm x$ and holding $\bm k$ fixed:\footnote{We could equally have elected to take a partial Fourier transform with respect to $t$ and hold $k^0$ fixed, but we will employ the former choice in the calculations to come.}
\begin{equation}
    P_\Omega \phi(t;{\bm k}) = \int_{|(k^0)^2 - {\bm k}^2| \leq \Omega^2} \frac{\dee k^0}{2\pi} \tilde \phi(k^0, {\bm k}) e^{-i k^0 t}
\end{equation}
A conventional sampling theorem then applies to $P_\Omega\phi(t; {\bm k})$ because the allowed frequencies, $k^0$, form a compact set.
In particular, notice that arbitrarily large magnitudes of ${\bm k}$ are still allowed, but the bandwidth in time for $P_\Omega\phi(t;{\bm k})$ falls to zero as $|{\bm k}| \rightarrow \infty$; the spatial modes ``freeze out'' and their density of degrees of freedom in time falls to zero.
Furthermore, these notions transform covariantly: a spatial mode that contracts under a Lorentz transformation acquires a smaller bandwith in time, which is consistent with the dilation of its degrees of freedom in time.
See Ref.~\cite{Kempf:2012sg} for a more extensive exposition.

Let us now return to a general Lorentzian manifold $\mathcal{M}$ and focus on the sharp covariant cutoff $P_\Omega$.
As a convenient piece of terminology, we will say that $P_\Omega \phi(x)$ is a \emph{covariantly bandlimited} field.
Conceptually, we implement the covariant cutoff at the level of the quantum field theoretic path integral by only integrating over covariantly bandlimited fields.
For example, consider the Feynman propagator, $G_F$, which can be written in terms of a path integral as
\begin{equation} \label{eq:GF}
    i G_F(x,x') = \frac{\int \mathcal{D}\phi ~ \phi(x) \phi(x') e^{iS[\phi]}}{\int \mathcal{D}\phi ~ e^{iS[\phi]}}.
\end{equation}
The covariantly bandlimited propagator, which we denote $G_F^\Omega$, is then given by
\begin{equation} \label{eq:cutoff-GF-PI}
    i G_F^\Omega(x,x') = \frac{\int_{B_\mathcal{M}(\Omega)} \mathcal{D}\phi ~ \phi(x) \phi(x') e^{iS[\phi]}}{\int_{B_\mathcal{M}(\Omega)} \mathcal{D}\phi ~ e^{iS[\phi]}},
\end{equation}
where $B_\mathcal{M}(\Omega) \equiv \mrm{span}\{\psi_\lambda \, | \, \Box \psi_\lambda = \lambda \psi_\lambda, |\lambda| \leq \Omega^2\}$ denotes the space of covariantly bandlimited fields on $\mathcal{M}$.

In practice, we implement the covariant cutoff using the projectors $P_\Omega$.
One can view the propagator $G_F(x,x')$ as the integral kernel of an operator on $L^2(\mathcal{M})$ that is the right inverse of the d'Alembert operator.
The bandlimited propagator is then obtained by projecting onto $B_\mathcal{M}(\Omega)$:
\begin{equation} \label{eq:GFc}
    G_F^\Omega = P_\Omega G_F P_\Omega
\end{equation}
This prescription is exactly equivalent to the path integral prescription \eqref{eq:cutoff-GF-PI} when the scalar field's action is of the form
\begin{equation} \label{eq:scalar-action}
    S[\phi] = \int \dee^{d+1}x \sqrt{-g} ~ \phi F(\Box) \phi,
\end{equation}
as shown in \App{app:PI-projector-equivalence}.
In particular, this includes the case of a free scalar field, with $F(\Box) = \Box - m^2$.

\subsubsection*{Cutoff fluctuation spectrum}

Our next goal is to impose a covariant cutoff on the fluctuation spectrum of a scalar field in a FLRW spacetime.
We consider FLRW spacetimes with no spatial curvature, for which we may write the line element
\begin{equation} \label{eq:FLRW-line-element}
    \dee s^2 = a^2(\eta) (-\dee \eta^2 + \dee x_i \dee x^i).
\end{equation}
We choose Cartesian spatial coordinates $x^i$, and for an inflating spacetime, the conformal time takes values $\eta \in (-\infty, 0)$.

The fluctuation spectrum of a scalar field on such a spacetime was defined in \Eq{eq:scalar-fluc-spec} in terms of its mode functions and in \Eq{eq:scalar-spec-from-2pt} in terms of its two-point function.
The two-point function coincides with the Feynman propagator at equal times, so let us write
\begin{equation}
\Delta^2_\phi(\eta,k) =  \left. 4\pi k^3 |G_F(\eta,\eta';k)| \, \right|_{\eta = \eta'} .
\end{equation}
$G_F(\eta,\eta';k)$ denotes the spatial Fourier transform of $G_F$ with respect to ${\bm x}$, and it depends only on the magnitude $k \equiv |{\bm k}|$ because of spherical symmetry.
This motivates us to define the (covariantly) cutoff fluctuation spectrum by
\begin{equation}
    (\Delta_\phi^\Omega)^2(\eta,k) =  \left. 4\pi k^3 |G_F^\Omega(\eta,\eta';k)| \, \right|_{\eta = \eta'},
\end{equation}
and so we must compute the cutoff propagator $G_F^\Omega$.

In fact, what we are more interested in is the correction to the fluctuation spectrum due to imposing a covariant cutoff.
When we impose the covariant cutoff, the Feynman propagator of course changes:
\begin{equation}
G_F ~ \rightarrow ~ G_F^\Omega \equiv G_F + \delta G_F
\end{equation}
The change in the fluctuation spectrum is then
\begin{equation} \label{eq:change-in-fspec-exact}
\delta \Delta_\phi^2 = 4\pi k^3 \left( |G_F + \delta G_F| - |G_F| \right).
\end{equation}
In practice, we will always have that $|\delta G_F| \ll |G_F|$, and so we can compute the difference \eqref{eq:change-in-fspec-exact} via a Taylor series.
One finds, to lowest order in $\Re~\delta G_F$ and $\Im~\delta G_F$, that the relative change in $\Delta_\phi^2$ is given by
\begin{equation}
\label{eq:reldiffscalar}
\frac{\delta \Delta_\phi^2}{\Delta_\phi^2} = \mrm{Re} \left( \frac{\delta G_F}{G_F} \right) + O(\delta G_F^2).
\end{equation}
In a similar vein, while the covariantly bandlimited propagator is formally given by \Eq{eq:GFc}, we can access the correction to it directly by rewriting \eqref{eq:GFc} in terms of the complementary projector $P_\Omega^\perp = I - P_\Omega$:
\begin{equation}
    G_F^\Omega = G_F + (P_\Omega^\perp G_F P_\Omega^\perp - P_\Omega^\perp G_F - G_F P_\Omega^\perp).
\end{equation}
The bracketed term is therefore the correction $\delta G_F$.

The calculation of $\delta G_F$ begins with writing down the Sturm-Liouville eigenvalue problem: $\Box u(\eta, {\bm x}) = \lambda u(\eta, {\bm x})$.
Notice, however, that a spatial Fourier transform with respect to ${\bm x}$ preserves the spectrum of the d'Alembertian, i.e. if $\Box u(\eta, {\bm x}) = \lambda u(\eta, {\bm x})$, then $\Box_{\bm k} u(\eta, {\bm k}) = \lambda u(\eta, {\bm k})$.
We may therefore impose the covariant cutoff on each spatial mode individually, which will be practical since we are ultimately interested in computing the correction to the spatial Fourier transform of the propagator on a mode-by-mode basis.
Next, we fix boundary conditions so that $\Box_{\bm k}$ is self-adjoint, and we identify its spectrum.
Then, with eigenfunctions and eigenvalues in hand, for each $k$ we construct the projector $P_\Omega^\perp$ and use it to obtain $\delta G_F$, and hence $\delta \Delta^2_\phi / \Delta^2_\phi$.

\subsubsection*{Example: de Sitter inflation}
\label{sec:dS-fluctuation-spectrum}

As a both productive and necessary example, let us explicitly compute $\delta G_F$ for a massless scalar field in de Sitter spacetime.
This amounts to the choice of scale factor $a(\eta) = (-H\eta)^{-1}$, and we also specialize to four spacetime dimensions from now on.
We will only describe the calculation in broad strokes here, leaving a complete account of all the details to \App{app:details}.
In summary, the calculation essentially proceeds in six steps:
\begin{enumerate}
    \item Starting with the de Sitter $k$-d'Alembertian $\Box_k$, we write down the two linearly independent solutions of the eigenvalue equation $\Box_k u = \lambda u$ (Eqs.~\eqref{eq:SL-J} and \eqref{eq:SL-Y}). One solution is normalizable in $L^2((-\infty,0), a^4(\eta)\, \dee \eta)$ when $\lambda < 9H^2/4$, and so self-adjoint realizations of $\Box_k$ have point spectrum in this range. In contrast, both solutions are non-normalizable for $\lambda \geq 9H^2/4$, and so self-adjoint realizations of $\Box_k$ have continuous spectrum in this range.
    \item We use an orthonormality relation (\Eq{eq:pt_spec_eigenf}) among point spectrum eigenfunctions, $\psi_n$, to determine the different possible point spectra corresponding to different self-adjoint realizations of $\Box_k$ as an operator on $L^2((-\infty,0), a^4(\eta)\, \dee \eta)$.
    \item We fix a particular choice of self-adjoint extension by requiring that $\Box_k (G_F^h \psi_n) = \psi_n$, where $G_F^h$ is the Hermitian part of the Feynman propagator (\Eq{eq:GF-herm}). The latter is calculated according to canonical quantization with the Bunch-Davies vacuum chosen as the field's vacuum state (\Eq{eq:GF-dS}). Choosing a different vacuum state would therefore pick out a different self-adjoint extension of $\Box_k$.
    \item We then determine the continuous spectrum eigenfunctions, $\varphi_q$ (\Eq{eq:cts_spec_eigenf}), by requiring that they be orthogonal to the $\psi_n$, as well as mutually continuum-normalized.
    \item Next, we construct the projector $P_\Omega^\perp$ (\Eq{eq:P-Omega-perp}), which we use to write down an expression for the change in the Feynman propagator, $\delta G_F$, due to the covariant cutoff.
    \item Finally, we argue that we can neglect the point spectrum contribution to $\delta G_F$, and we make several useful approximations for the continuous spectrum contribution to arrive at a compact final expression for $\delta \Delta^2_\phi/ \Delta^2_\phi$.
\end{enumerate}

We ultimately find that
\begin{equation}\label{eq:final_answer}
    \frac{\delta \Delta^2_\phi}{\Delta^2_\phi} \approx \frac{ 2 I(Q,x) \left[ \left(\tfrac{2}{\pi}\right)^{3/2} Y_{3/2}(x) + \tfrac{4}{\pi^3}  I(Q,x) \right] }{J_{3/2}(x)^2 +Y_{3/2}(x)^2},
\end{equation}
where $I$ is the integral 
\begin{equation} \label{eq:I-integral-app-mainbody}
    I(Q,x) = - \int_0^\infty \dee b ~ e^{w(Q,b;2/(x e))} \sin\left( W[Q,b;2/(x e)] \right)
\end{equation}
and where we have defined $x = -k\eta$, $Q = \sqrt{\sigma^{-2}-9/4}$, and $\sigma = H/\Omega$.
The functions $w$ and $W$ are given by
\begin{align}
    w(Q,b;2/(xe)) &= - \frac{1}{2} \left(b + \frac{3}{2}\right) \ln\left(Q^2+b^2\right) - Q \arctan\left(\frac{b}{Q}\right) - b \ln\left(\frac{2}{xe}\right), \\[2mm]
    W(Q,b;2/(xe)) &= \frac{Q}{2} \ln\left(Q^2+b^2\right) - \left(b + \frac{3}{2}\right) \arctan\left(\frac{b}{Q}\right) + Q \ln\left(\frac{2}{xe}\right) .
\end{align}
\Eq{eq:final_answer} makes manifest that $\delta \Delta^2_\phi / \Delta^2_\phi$ is only a function of two independent parameters: $x$, which characterizes when we evaluate the fluctuation spectrum relative to horizon crossing, and $\sigma$, the ratio of the cutoff and Hubble scales.

Although the integral \eqref{eq:I-integral-app-mainbody} cannot be evaluated in closed form, note that for $\sigma\ll 1$, i.e. $Q\gg1$, the exponential function in the integrand decays much more rapidly than the sine function oscillates. Hence, in this limit, the integral can be well-approximated by expanding the functions $w$ and $W$ up to second order in $b$ and analytically evaluating the resulting Gaussian integral. This results in the asymptotic expansion\footnote{The next term in the expansion for $I$ is 
\begin{align}
    \frac{\left(3 \ln (2Q/x)+2\right) \cos \left(Q-Q \ln(2Q/x)\right)}{2 Q^{5/2} \ln ^3(2 Q/x)},
\end{align}
which is smaller than the leading term by a factor of $1/Q$. Hence for $Q\sim 1/\sigma\sim 10^5$, the leading approximation is quite accurate.
}
\begin{align}\label{eq:I-approx}
    I \sim \frac{\sin \left(Q-Q \ln(2Q/x)\right)}{Q^{3/2} \ln(2Q/x)}.
\end{align}
We see that $I\ll 1$ for $Q\gg 1$, and thus to a good approximation we can further neglect the $I^2$ term in the expression \eqref{eq:final_answer} for the relative change in $\Delta_\phi$, which gives
\begin{equation} \label{eq:final_approx}
    \frac{\delta \Delta^2_\phi}{\Delta^2_\phi}
    \approx
    -\frac{4(\cos x + x \sin x)}{\pi(1+x^2)}
    \frac{\sin (Q-Q\ln[2Q/x])}{Q^{3/2} \ln[2Q/x]}.
\end{equation}

\begin{figure}[t]
    \centering
    \includegraphics[scale=0.75]{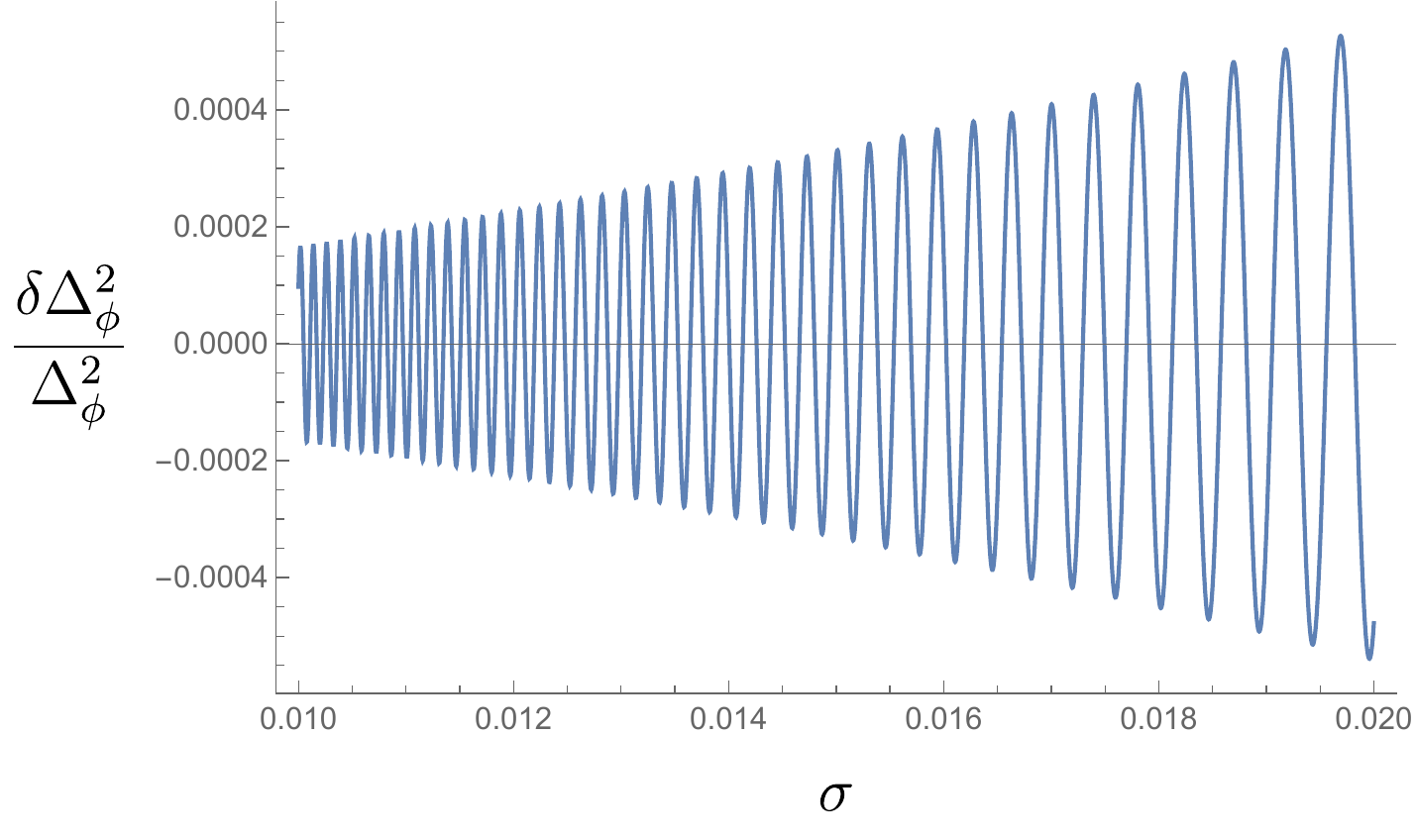}
    \caption{Relative change in $\Delta^2_\phi$ due to a sharp covariant cutoff as a function of $\sigma = H/\Omega$ with $x=1$ fixed to horizon crossing. For the purposes of clearly illustrating the oscillations, here we plot $\delta \Delta_\phi^2/\Delta_\phi^2$ for a fairly large value of $\sigma \sim 10^{-2}$.}
    \label{fig:dS-correction}
\end{figure}

It is expected that inflationary fluctuations freeze out (i.e. stop fluctuating) as they cross the horizon, so from now on we will set $x=1$ to reflect this expectation.
Then, if we also approximate $Q \approx \sigma^{-1}$, we arrive at the following compact expression for the relative change in $\Delta_\phi$:
\begin{equation}\label{eq:final_answer-approximation}
    \left. \frac{\delta \Delta^2_\phi}{\Delta^2_\phi} \right|_{x=1}
    \approx  \frac{2(\cos1+\sin1)}{\pi}
    \frac{\sigma^{3/2}}{\ln (\sigma/2)}
    \sin (\omega(\sigma)\sigma).
\end{equation}
In the equation above, we defined the $\sigma$-dependent frequency
\begin{align}
    \omega(\sigma)\equiv  
    \frac{1}{\sigma^2}\ln\frac{\sigma e}{2}.
\end{align}
These oscillations are a key feature of our predicted correction to the power spectrum.
A plot of the correction to $\Delta_\phi^2$ as a function of $\sigma$, in which these characteristic oscillations can be seen, is shown in \Fig{fig:dS-correction}.

\subsubsection*{Predictions for slow-roll FLRW inflation}

For the purpose of comparing with cosmological data, given a slowly-rolling inflationary spacetime, the goal is to produce a prediction for $\delta \Delta^2_\phi/\Delta^2_\phi$ evaluated at horizon crossing as a function of the comoving mode $k$, namely
\begin{equation} \label{eq:prediction}
    \left. \frac{\delta \Delta^2_\phi}{\Delta^2_\phi} \right|_{aH = k}.
\end{equation}
For a de Sitter scale factor, horizon crossing happens at $x = - k \eta = 1$ for all modes, and so \eqref{eq:prediction} is a constant correction for all $k$.
This is just a reflection of the fact that the proper size of the de Sitter horizon is constant in time, and so all modes are the same proper size when they cross the horizon.
Consequently, the magnitude of their fluctuations is the same, as well as the magnitude of the correction due to a covariant cutoff.\footnote{Explicitly, here both $x = -k\eta$ and $\sigma = H/\Omega$ are constant in using \Eq{eq:final_answer} to compute \eqref{eq:prediction}.}
On the other hand, in a slowly-rolling inflationary spacetime, the proper size of the cosmological horizon is slowly changing, and so different modes will have different proper wavelengths when they cross the horizon. 
The prediction \eqref{eq:prediction} will therefore be nontrivial as a function of $k$.

The calculation that we carried out for de Sitter in \Sec{sec:dS-fluctuation-spectrum} is intractable for a generic FLRW spacetime, including even for a simple example such as a power law scale factor. 
Our strategy will therefore be to approximate the prediction \eqref{eq:prediction} for a given slowly-rolling FLRW spacetime by a succession of instantaneously de Sitter calculations.
That is, we will use the de Sitter result \eqref{eq:final_answer} to compute \eqref{eq:prediction}, but with a Hubble parameter $H$ that depends on $k$.
Intuitively, such an adiabatic approximation will be accurate provided that the true time-dependent Hubble parameter of the FLRW spacetime, $H(\eta)$, evolves sufficiently slowly, which is what we expect during slow-roll inflation.
We discuss this adiabatic, or ``slow-roll'' approximation and additional supporting evidence for its validity in \App{sec:approximation}.

In practice, for each given comoving mode $k$, we must fix the values of $\eta$ and $H$, or equivalently the values of $x$ and $\sigma$, in \Eq{eq:final_answer-approximation}.
Since we are interested in the correction to a mode's fluctuations when it crosses the cosmological horizon, and since \Eq{eq:final_answer-approximation} was derived for a de Sitter background, we will set $\eta$ to be the de Sitter horizon-crossing time; that is, we fix $x = -k\eta = 1$ in \Eq{eq:final_answer-approximation}.
Then, to fix the value of $H$, we use the horizon-crossing condition $a(\eta) H(\eta) = k$ for the slowly-rolling $a(\eta)$ to determine the value of $H(\eta)$ at horizon crossing.
We show this calculation explicitly in the next section.

\section{Corrections to primordial power spectra}
\label{sec:prediction}

In this section, we compute the correction to the primordial power spectrum due to a covariant natural UV cutoff using realistic cosmological parameters.
We focus on the scalar spectrum, $\Delta_\mathcal{R}^2$, but the calculation for the tensor spectrum, $\Delta_\mathcal{T}^2$, is completely analogous.

As input to the prediction \eqref{eq:prediction}, we need the Hubble parameter seen by a mode $k$ when it crosses the cosmological horizon.
We compute this by comparing the theoretical form of $\Delta_\mathcal{R}^2$ (without cutoff) given by \Eq{eq:scalar-spectrum} with its observational parameterization \eqref{eq:scalar-spectrum-pheno}.\footnote{Note that we are justified in using the uncorrected value of $\Delta_\mathcal{R}^2$ to compute $H$ because the error incurred in doing so is of second order, whereas we are computing a first order correction.}
Equating the two, we obtain
\begin{equation} \label{eq:Hubble-eff}
    \left. \frac{H^2}{\epsilon} \right|_{aH = k} = \Mpl^2 \pi A_s \left( \frac{k}{k_\star}  \right)^{n_s-1}.
\end{equation}
In order to get $H^2$ as a function of $k$, we need to account for how the first slow-roll parameter $\epsilon$ depends on $k$.
To leading order in the slow-roll expansion, this is given by
\begin{align}
    \frac{d\ln\epsilon}{d\ln k} = 
    2(\epsilon+\delta),
\end{align}
where $\delta$ is the second slow-roll parameter, as defined in \Eq{eq:slow_roll_params} \cite{dodelson2003modern}.
Taylor expanding $\ln \epsilon(k)$ about the pivot scale $k_\star$ therefore yields
\begin{align}
    \epsilon(k)=\epsilon_\star\left(\frac{k}{k_\star}\right)^{2(\epsilon_\star+\delta_\star)},
\end{align}
where $\epsilon(k_\star)\equiv \epsilon_\star$, $\delta(k_\star)\equiv \delta_\star$, and we suppress terms that are quadratic or higher in $\epsilon_\star$ and $\delta_\star$.
Similarly, to leading order in the slow-roll expansion, the spectral tilt is given by
\begin{align}
    n_s = 1-4\epsilon_\star -2\delta_\star.
\end{align}
Dividing by the cutoff scale $\Omega$, we therefore find that the ratio of the Hubble and cutoff scales at the time when the mode $k$ crosses the horizon is given by
\begin{align}\label{eq:sigma}
    \sigma(k) \equiv \left. \frac{H}{\Omega} \right|_{aH=k} =  \frac{\Mpl}{\Omega}\sqrt{\pi A_s\epsilon_\star}\left(\frac{k}{k_\star}\right)^{-\epsilon_\star}.
\end{align}

From the expression above, we see that $\sigma(k)$ depends on two cosmological parameters, $A_s$ and $\epsilon_\star$, as well as single new parameter, namely, the covariant cutoff scale $\Omega$.
Finally, we can use \Eq{eq:sigma} in \Eq{eq:final_answer} or \Eq{eq:final_answer-approximation} to obtain a prediction for the corrections $\delta \Delta_\mathcal{R}^2/\Delta_\mathcal{R}^2$ to the scalar primordial power spectrum, where we take the placeholder scalar field $\phi$ to be the Mukhanov-Sasaki variable, $\mathcal{R}$.
The corrections are given by
\begin{align}\label{eq:main_prediction}
    \left. \frac{\delta \Delta^2_\calR}{\Delta^2_\calR}\right|_{aH=k}
    =
    \mathcal{C}
    \frac{\sigma(k)^{3/2}}{\ln(\sigma(k)/2)} \sin\left(\omega(k)\,  \sigma(k)\right),
\end{align} 
where $\mathcal{C} = 2(\cos 1 + \sin 1)/\pi$,
\begin{align}
    \omega(k) \equiv \omega(\sigma(k))
    =  \frac{1}{\sigma(k)^2}\ln\frac{e\sigma(k)}{2},
    \label{eq:oscillation-frequency}
\end{align}
and where $\sigma(k)$ is given by \Eq{eq:sigma}.
Both the amplitude and frequency of the oscillations track the ratio $\sigma(k)$ of the Hubble and cutoff scales at horizon crossing, and as a result both are functions of $k$. 

Computing a concrete correction using \Eq{eq:main_prediction} requires specifying values for $A_s$, $\epsilon_\star$, and $\Omega$.
The amplitude $A_s$ is tightly constrained by CMB observations \cite{Planck:2018jri} and takes the value $A_s \approx 2.1\times 10^{-9}$.
Assuming that the Hubble length increases during inflation, as it must for inflation to ultimately end, $\epsilon_\star$ must be positive. Additionally, CMB data gives an upper bound on $\epsilon_\star$ at around 0.004 for the pivot scale $k_\star = 0.05 \text{ Mpc}^{-1}$ \cite{Planck:2018jri}. Thus, 
\begin{align}
    0\le\epsilon_\star\lesssim 0.004.
\end{align}
While $\Omega$ is a new parameter that is to be determined experimentally, it is already subject to certain constraints out of principle.
First, $\Omega$ is presumably bounded from above by the Planck scale $\Mpl$.
There should also be a lower bound on $\Omega$ to ensure that $\sigma(k)\le 1$ for all momenta $k$ corresponding to modes which we observe in the CMB, i.e. for $k_\text{min}<k<k_\text{max}$, where $k_\text{min} = 10^{-4} \text{ Mpc}^{-1}$ and $k_\text{max} = 10 \text{ Mpc}^{-1}$. This ensures that the cutoff energy is above the Hubble energy for all visible modes, which should be true because effective field theory (which should be valid only below the scale of quantum gravity) seems to make accurate predictions for inflationary perturbations.
Because $\epsilon_\star>0$, $\sigma(k)$ is a decreasing function of $k$, and so $\sigma(k)\le 1$ is equivalent to $\sigma(k_\text{min})\le 1$.
Hence, using Eq. \eqref{eq:sigma}, we expect that
\begin{align} \label{eq:Omega-inequality}
    \sqrt{\pi A_s \epsilon_\star}\left(\frac{k_\text{min}}{k_\star}\right)^{-\epsilon_\star}
    \le \frac{\Omega}{\Mpl} \le 1.
\end{align}

\begin{figure}[ht!]
    \centering
    \includegraphics[width=0.53\textwidth]{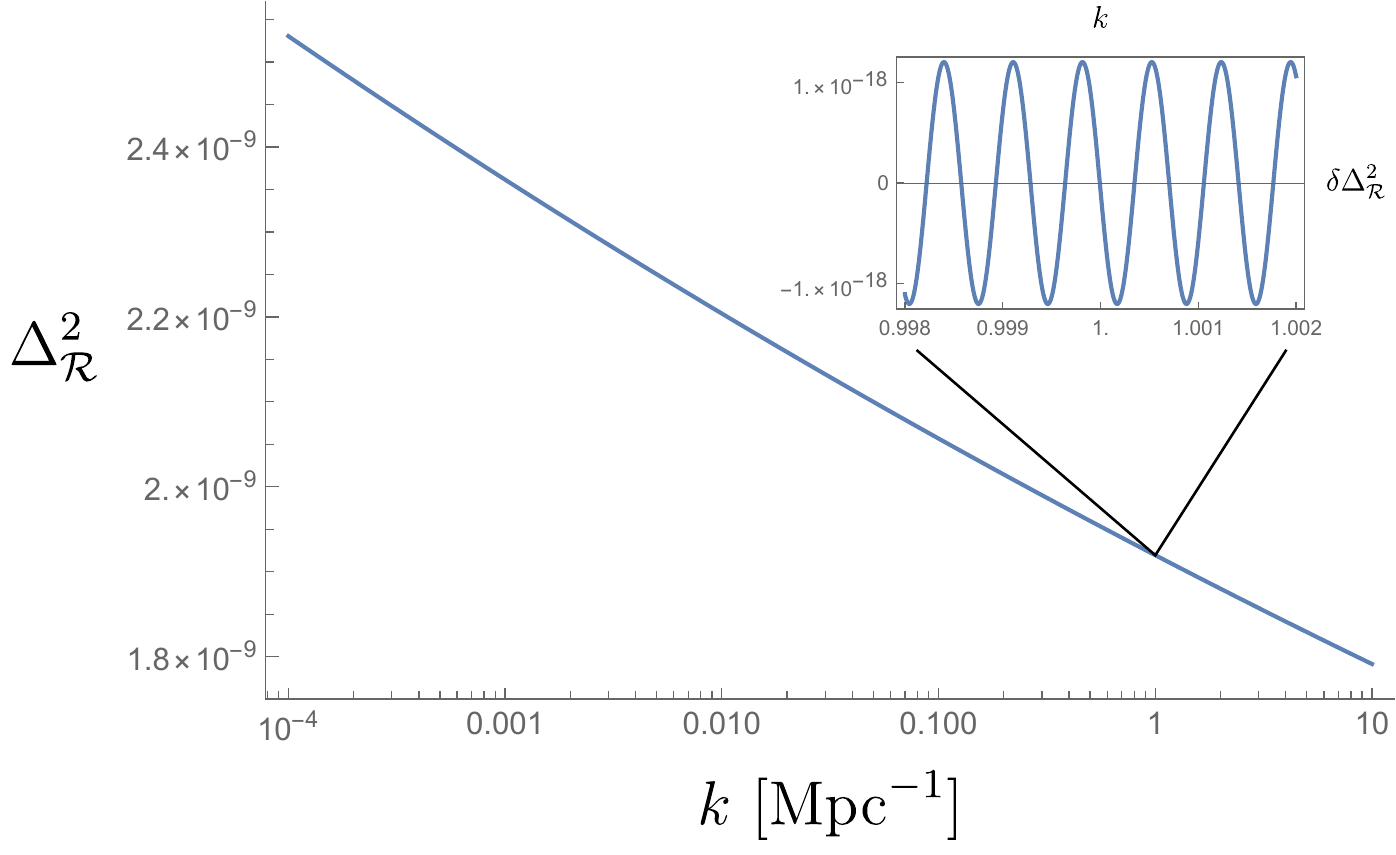} \\ (a) \\
    \includegraphics[width=0.53\textwidth]{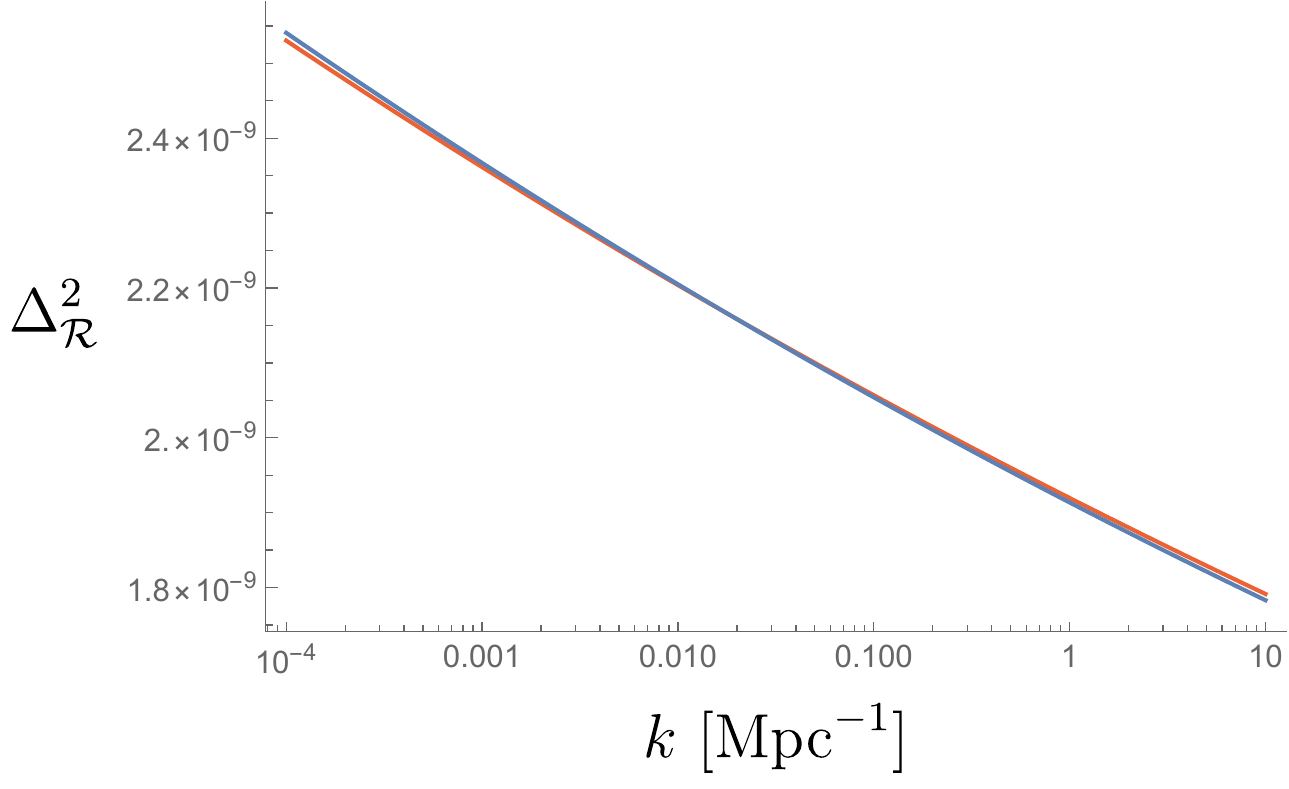} \\ (b) \\
    \includegraphics[width=0.53\textwidth]{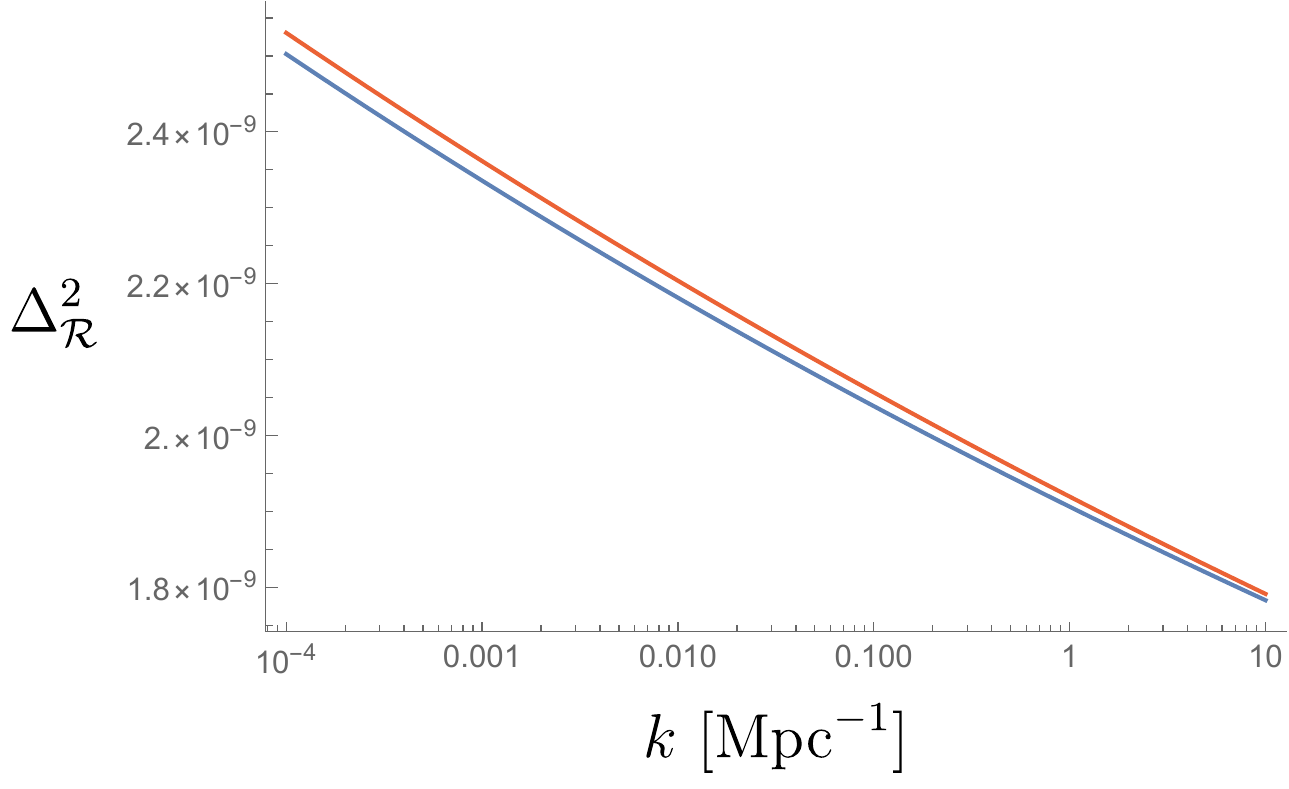} \\ (c)
    \caption{(a) The primordial scalar power spectrum, corrected due to a sharp cutoff $\Mpl / \Omega = 1$, with $\epsilon_\star = 0.003$. 
    The inset plot shows the correction itself, $\delta \Delta_\calR^2$, over a small interval of modes which resolves the oscillations.
    (b) The uncorrected primordial power spectrum is shown in red, and the corrected spectrum due to a sharp cutoff $\Mpl / \Omega = 2.1 \times 10^4 $ is shown in blue, also with $\epsilon_\star = 0.003$.
    (c) The same curves are plotted as in (b), but here with $\Mpl / \Omega = 2.1 \times 10^4 $ and $\epsilon_\star = 0.004$.
    In all plots, other parameter values are $A_s = 2.1 \times 10^{-9}$, $k_\star = 0.05~\mrm{Mpc}^{-1}$, and $n_s = 0.97$.}
    \label{fig:prediction}
\end{figure}

Given these constraints, the model \eqref{eq:main_prediction} should be fit to observations of the primordial power spectrum, which can be inferred from late time power spectra, such as those of the CMB. We leave a detailed analysis of this fitting to future work. Here, to get a qualitative sense of what our prediction looks like, in Fig.~\ref{fig:prediction} we plot the baseline PPS and its correction for particular choices of $\epsilon_\star$ and $\Omega$. We find that a sharp covariant natural UV cutoff causes small $k$-dependent oscillations which are superimposed on the non-cutoff primordial power spectrum.

First we note that the oscillation frequency $\omega(\sigma(k))$ is the most robust predicted feature of covariant natural UV cutoffs.
So far, we have focused on a sharp cutoff $P_\Omega = \theta(\Omega^2 - |\Box|)$; however, the class of cutoffs specified by \Eq{eq:cutoff} that we could consider is much larger.
For example, we could consider a cutoff $f(\Box)$ that smooths out the Heaviside step function in $P_\Omega$.
Doing so effectively introduces a free functional parameter (which characterizes the degree of smoothness), yet even so, the oscillation frequency is largely independent of such precise details of the cutoff.
This is in contrast to the oscillations' amplitude and phase, which depend more strongly on such details.
These points are discussed more thoroughly in \App{sec:detailed-features}.
In a nutshell, smoothing out the cutoff shifts the oscillations' phase and tends to reduce their amplitude.
The former has little practical consequences, but the latter makes the oscillations harder to observe.
With these findings in mind, here we continue to focus on the case of a sharp cutoff, with the understanding that this corresponds to the strongest possible signal.

\begin{figure}
    \centering
    \includegraphics[width=0.48\linewidth]{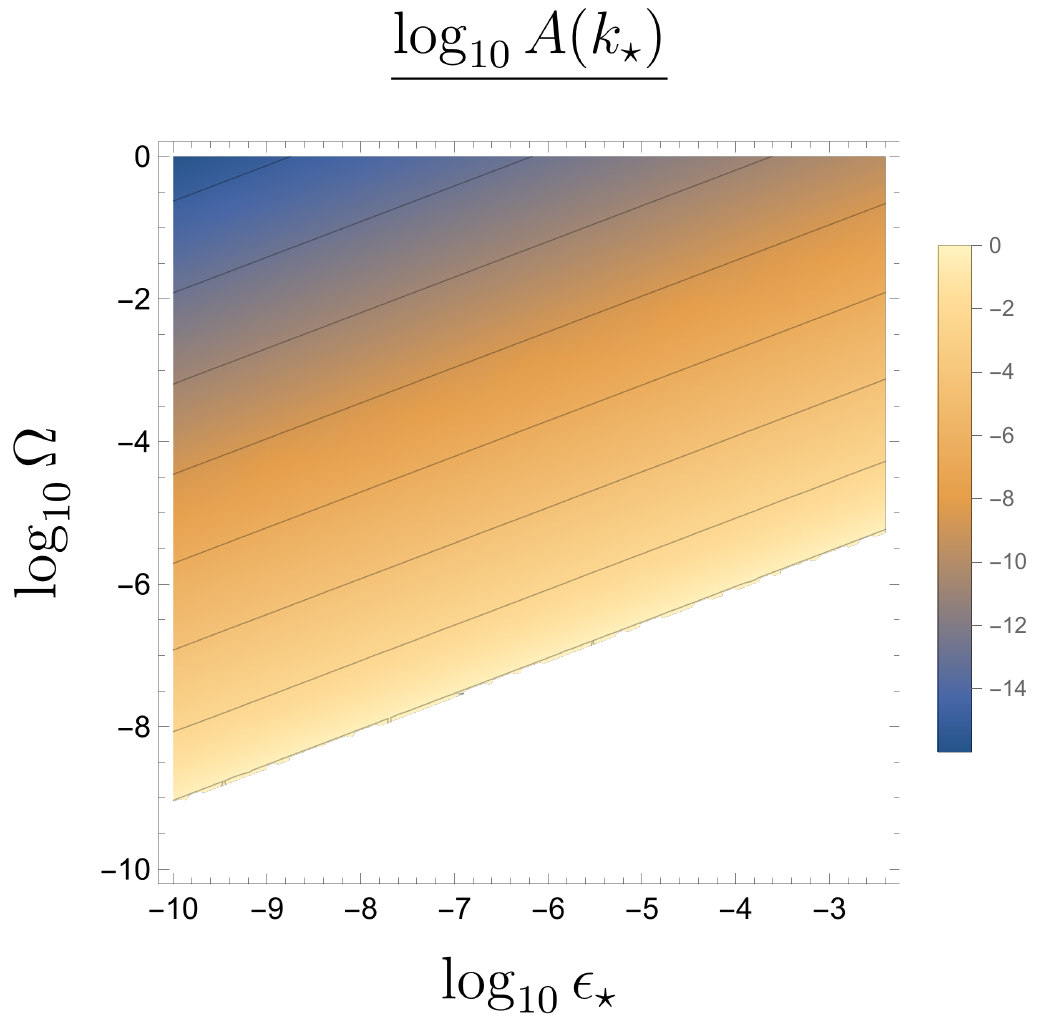} ~ 
    \includegraphics[width=0.48\linewidth]{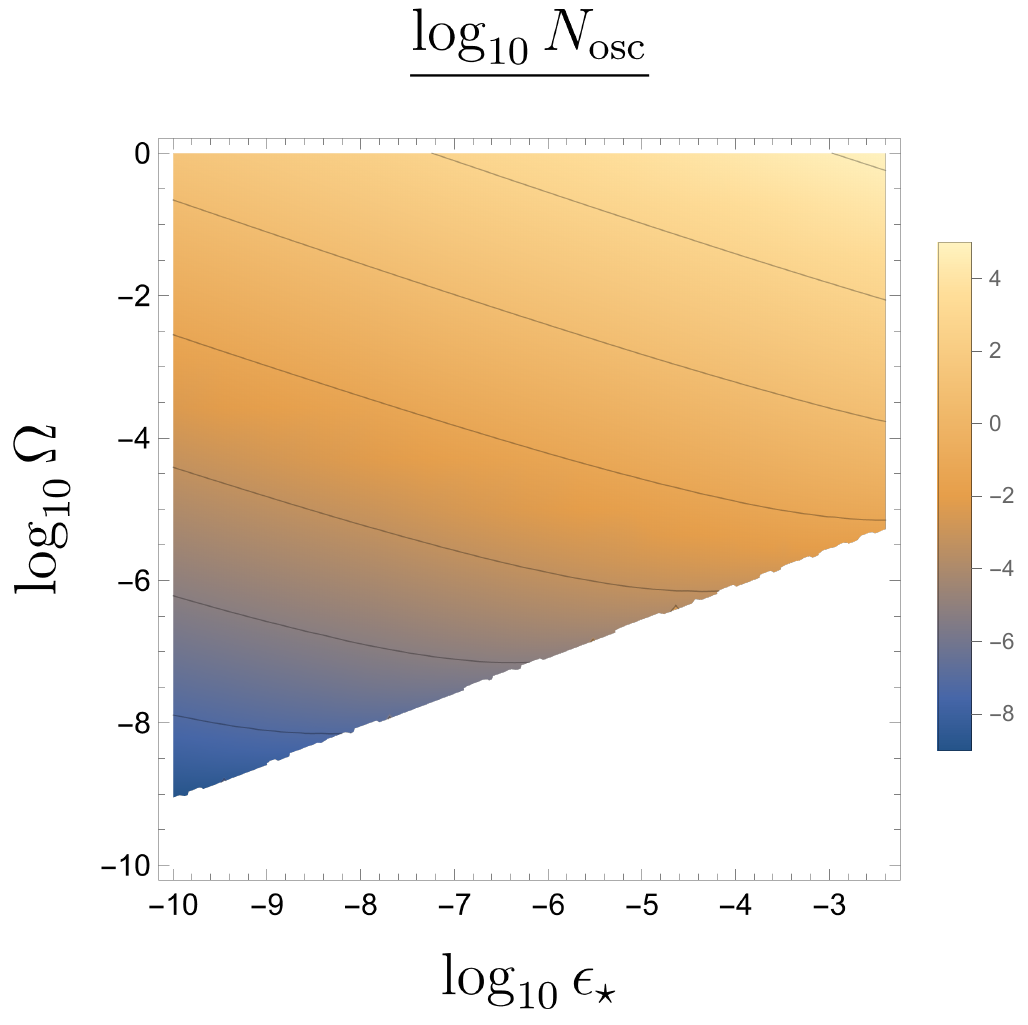} \\
    ~ (a) \hspace{0.48\linewidth} (b)
    \caption{Density plots illustrating how (a) the amplitude, $A$, and (b) the number of oscillations in the visible window of modes, $N_\mrm{osc}$, scale with the parameters $\epsilon_\star$ and $\Omega$. The white regions correspond to disallowed parameter values for which $\sigma = H/\Omega > 1$.}
    \label{fig:Amp_and_Nosc}
\end{figure}

Next, we would like to get a sense for how the predicted signal depends on $\Omega$ and $\epsilon_\star$, which are the least constrained parameters of the model.
To this end, define the amplitude and number of oscillations in the visible window $k_\text{min}<k<k_\text{max}$ as\begin{align}
    A(k) &\equiv \frac{\sigma(k)^{3/2}}{\ln(\sigma(k)/2)}, \\[2mm]
    N_\text{osc} &\equiv \frac{\left|\omega(k_\text{max})\sigma(k_\text{max})-\omega(k_\text{min})\sigma(k_\text{min})\right|}{2\pi}.
\end{align}
We plot $A(k_\star)$ and $N_\text{osc}$ as functions of $\epsilon_\star$ and $\Omega$ in Fig. \ref{fig:Amp_and_Nosc}. For concreteness we plot the amplitude at the pivot scale, but because the amplitude changes relatively little with $k$, the plot would look qualitatively the same had we chosen any other scale between $k_\text{min}$ and $k_\text{max}$.

Let us consider the scaling of $A(k)$ and $N_\text{osc}$ with the parameters $\Omega$ and $\epsilon_\star$. Since $\sigma\sim \epsilon_\star^{1/2} \Mpl /\Omega$, ignoring logarithmic corrections we see that
\begin{align}
    A&\sim \left(\frac{\Mpl}{\Omega}\right)^{3/2}\epsilon_\star^{3/4},\\[2mm]
    N_\text{osc} &\sim \frac{\Omega}{\Mpl} \epsilon_\star^{1/2}.
\end{align}
First, note that as the cutoff $\Omega$ increases, the predicted amplitude of corrections to the PPS decreases.
This makes sense, because the correction should vanish in the limit that the cutoff is removed by taking $\Omega \rightarrow \infty$.
Interestingly, the scaling $\Omega^{-3/2}$ of the amplitude exactly interpolates between previous findings in the literature that the decay should be linear \cite{Easther:2001fi,Easther:2005yr} or quadratic \cite{Kempf:2001fa,Frob:2012ui,Kaloper:2002uj} in the ratio of cutoff and Hubble scales.
That $N_\text{osc}$ increases with increasing $\Omega$ is consistent with the intuition that the oscillations arise as a sort of ``interference'' between the Hubble and cutoff scales \cite{Chatwin-Davies:2016byj}. Heuristically, peaks in $\delta \Delta_\calR^2(k)$ occur for $k$ such that an integer number of cutoffs lengths fit into $1/H(k)$, with similar reasoning applying to troughs.
Increasing $\Omega$ therefore accelerates this interference pattern when scanning through a range of modes $k$.

We can understand the dependence of $A$ on $\epsilon_\star$ by noting that, from the perspective of the uncorrected PPS \eqref{eq:scalar-spectrum}, increasing $\epsilon_\star$ may be compensated for by increasing $H$, which brings the Hubble scale closer to the cutoff scale.
To understand how $N_\text{osc}$ scales with $\epsilon_\star$, note that taking $\epsilon_\star \rightarrow 0$ corresponds (to leading order) to a static Hubble parameter for which the correction $\delta \Delta_\calR^2(k)$ is just a constant offset with no oscillations.
However, the $\epsilon_\star \rightarrow 0$ limit should be taken with a grain of salt in the leading order calculations here, since the Mukhanov-Sasaki variable is undefined if $\epsilon \rightarrow 0$, and $\Delta_\calR^2$ diverges.

The prediction for the primordial tensor power spectrum is computed similarly to the scalar case.
The only difference is that we write $\sigma$ in terms of the tensor parameters as
\begin{equation}
    \left. \sigma \right|_{aH=k} = \frac{ \sqrt{\pi A_t}}{4\Omega} \left(\frac{k}{k_\star}\right)^{n_t}.
\end{equation}
Notice however, that unlike in the case of scalar perturbations, the slow roll parameter $\epsilon_\star$ does not appear in this expression. Instead it is replaced by the tensor spectral tilt index $n_t$.
Should eventual measurement of the tensor spectrum be sensitive enough to determine a value for $n_t$, then $\Omega$ would be the only parameter left to constrain in the tensor spectrum.
Additionally, because the ratio of tensor to scalar amplitudes, $A_t/A_s$, is equal to $16\epsilon_\star$, a measurement of the tensor spectrum would also give a value to $\epsilon_\star$, in which case $\Omega$ would be the only parameter left to constrain in the scalar PPS as well.

\section{Discussion}
\label{sec:discussion}

\subsubsection*{Summary}

We calculated the signature that a generic, quantum gravity-motivated, natural UV cutoff would leave in primordial inflationary power spectra.

The UV cutoff that we considered takes the form of a cutoff on the spectrum of a scalar field's d'Alembertian, i.e., it is a large eigenvalue cutoff for scalar fields.
As such, it covariantly generalizes the notion of a maximum Fourier frequency to arbitrary Lorentzian manifolds.
It also admits a natural information theoretic interpretation as a cutoff on the density of field degrees of freedom in spacetime, in the sense of Shannon sampling theory; see \cite{Kempf:2012sg,Chatwin-Davies:2016byj}.

We implemented the natural UV cutoff in the language of path integrals by restricting the space of fields integrated over in a path integral to only those fields that are spanned by eigenfunctions of the d'Alembertian whose eigenvalues are less than the cutoff scale.
Conceptually, this can be thought of as discarding, in a covariant way, the contributions to the path integral of field configurations which fluctuate too far off shell.
In practice, this is equivalent to constructing projectors and using them to restrict operators to the subspace defined by the cutoff.
We illustrated this process by calculating the covariantly cut-off Feynman propagator for a massless scalar field in de Sitter spacetime and we explained how to generalize the result to slowly-rolling FLRW spacetimes in \Sec{sec:covariant-cutoff}.

Furthermore, we used the fact that the primordial scalar perturbation and each polarization of the tensor perturbation are massless scalar fields that propagate on a slowly-rolling FLRW background, and that the power spectra of their fluctuations are straightforward to calculate in terms of the Feynman propagator.
This allowed us to use our results for the covariantly bandlimited Feynman propagator to compute the correction that a covariant natural ultraviolet cutoff produces on the primordial power spectra.

While our calculations for the tensor and scalar spectra are analogous, we focused on the scalar spectrum, $\Delta^2_\mathcal{R}(k)$, due to its better experimental prospects.  
We found that the correction induced by the cutoff, $\delta \Delta^2_\mathcal{R}(k)$, takes the form of small $k$-dependent oscillations;\footnote{Log-oscillations, i.e., corrections to the primordial power spectrum that schematically oscillate like $\sin(\omega \ln k)$, are not unique to our prediction; see, e.g., \Ref{Easther:2005yr}, or \Ref{Calcagni:2016ofu} and Refs.~[27-43] therein, and also Refs.~\cite{Ashtekar:2021izi,Agullo:2021oqk} for oscillatory predictions from Loop Quantum Cosmology. To our knowledge, the chirping $k$-dependence of $\omega(k)$ in \Eq{eq:oscillation-frequency}, as well as the overall amplitude in \Eq{eq:main_prediction} for the case of a sharp cutoff, are specific to our prediction.} see Eqs.~\eqref{eq:main_prediction} and \eqref{eq:oscillation-frequency}.
The amplitude and phase of the oscillations depend moderately on the precise details of the cutoff, namely, how smoothly and over how many Planck lengths it turns on.
However, the oscillation frequency is a particularly robust characteristic of the prediction.
The frequency, as a function of $k$, tracks the ratio of the cutoff and Hubble scales when the mode $k$ crossed the Hubble length during inflation, and its functional form is completely fixed according to \Eq{eq:oscillation-frequency}.
The only free parameters of this prediction are the first slow-roll parameter $\epsilon_\star$, and the location of the cutoff; for fixed $\epsilon_\star$, the oscillation frequency decreases with decreasing $\Omega$. 
Therefore, in the case that the cutoff scale is located before the Planck scale, the predicted signature is stronger due to a larger amplitude and lower frequency.
This scenario could correspond to quantum gravitational effects becoming important at, e.g., the string scale.

\subsubsection*{Effective field theory of inflation}

Effective field theory has been used to systematically study the possible high energy corrections to inflation \cite{Cheung:2007st}. As always in effective field theory, one considers a Lagrangian with all possible local interaction terms that satisfy the appropriate symmetries, and one provides a Wilsonian cutoff below which the coupling coefficients to these terms can be deduced from experiments. By integrating out higher energy contributions to the path integral, the Wilsonian cutoff is lowered and the coupling coefficients flow to different values. Of course, in the absence of experiments, effective field theory does not pick out a particular theory (i.e. a particular set of coupling coefficients) as the correct one. 

What we are doing in this paper, on the other hand, is to study the free theory of inflaton and linearized gravitational perturbations, with a covariant cutoff imposed on the path integral. Unlike in the effective field theory framework, we are studying, therefore, one particular theory; there are no undetermined coefficients to fix using experiments. In particular, we emphasize that we did \textit{not} obtain our field theory by integrating out trans-Planckian contributions to the path integral.
Instead, we discard these contributions from the quantum field theoretic path integral and study the resulting modified theory.

Nevertheless, the formalism of effective field theory may well be broad enough to encompass also the UV modification to quantum field theory that we are considering here. 
To this end, a possible starting point is as follows.
A path integration over the space $B_\mathcal{M}(\Omega)$ of covariantly bandlimited fields can be re-written as a path integral over the space of all fields with an indicator function weighting the integration.
Explicitly,
\begin{equation}
    \int_{B_\mathcal{M}(\Omega)} \mathcal{D}\phi~e^{iS[\phi]} = \int \mathcal{D}\phi ~ \Pi_\Omega[\phi] e^{iS[\phi]},
\end{equation}
where the indicator function $\Pi_\Omega[\phi] = 1$ if $\phi \in B_{\mathcal{M}}(\Omega)$ and vanishes otherwise, or is a more complicated functional in the case of a soft cutoff.
We may then  exponentiate the indicator to formally obtain
\begin{equation}
    \int \mathcal{D}\phi~e^{i(S[\phi] - i \ln \Pi_\Omega[\phi])}.
\end{equation}
One could then attempt to understand $i \ln \Pi_\Omega[\phi]$ as higher-order corrections to the free field action $S[\phi]$.
We leave these considerations to future work.

\subsection*{Choice of vacuum state}

For our analysis, we chose the state of the quantum field to be the Bunch-Davies state in order to deviate as minimally as possible from the standard inflationary calculation.
To put our motivation another way, because we are modelling quantum gravitational physics by effectively discarding certain modes' contributions to the path integral, we did not want to conflate our modelling of Planck-scale physics with effects that arise from altering the vacuum states for these modes.
Moreover, there is evidence that deviations from the Bunch-Davies vacuum might interfere with gauge invariance and create divergences in perturbation theory \cite{Urakawa:2010kr}.  

That said, given suitable motivation, it is certainly possible to make a prediction, per our formalism, for a different choice of vacuum state.
Within our framework, changing the vacuum state can be accomplished by changing the choice of self-adjoint extension of the d'Alembertian (Step 3 of the calculation in \Sec{sec:covariant-cutoff}).
Such a change would eventually percolate to a different prediction for $\delta \Delta_\mathcal{R}^2$, which could be quite significant \cite{deBlas:2016puz,Navascues:2021qcp}, given that a different choice of vacuum state can significantly alter the uncorrected primordial power spectrum itself.

\subsubsection*{Power at large comoving scales}

In principle, our predictions are compatible with an apparent lack of power at large scales in measured CMB power spectra; see, e.g., Sec.~III.2.6 of Ref.~\cite{Perivolaropoulos:2021jda}.
A covariant cutoff at a sufficiently low energy scale could produce a long wavelength oscillation that amounts to a trough at large comoving scales that would reduce the power in the primordial spectrum there.
Such a possibility is illustrated in Fig.~\ref{fig:prediction}c; whether or not this suppression happens depends also on the value of the slow roll parameter $\epsilon_\star$. 
Should this tension be confirmed to be more than a statistical anomaly, it may help to further constrain the fit of our predicted template to observed data.

\subsubsection*{Observational prospects}
A natural covariant UV cutoff, $\Omega$, is widely expected to exist somewhere between the Hubble scale during inflation and the Planck scale.
$\Omega$ should be no larger than the Planck scale, because this is the scale at which we expect quantum gravitational effects to dominate, and hence the low-energy path integral description to break down. Also, $\Omega$ is expected to be not too close to the Hubble scale during inflation, in order to be consistent with the standard quantum field theoretic description of cosmological perturbations, which is supported by observations.
These expectations are encapsulated by the inequality \eqref{eq:Omega-inequality}.

Here, we calculated concrete predictions for the corrections to the primordial scalar and tensor power spectra that arise from such a cutoff, as a function of $\Omega$. The main prediction is that the presence of this UV cutoff results in oscillations on top of the familiar, nearly-scale-invariant, primordial power spectrum curve; see Fig. \ref{fig:prediction}. 

If the natural UV cutoff is located at what is presumably the upper limit of its possible range, i.e., at the Planck scale with $\Omega = \Mpl$, then, as Fig.~\ref{fig:prediction}a shows for a sharp cutoff and a representative value of $\epsilon_\star$, the predicted oscillations have an amplitude which is very small (about 9 orders of magnitude smaller than the mean value of the power spectrum) and their frequency is very high (there are roughly $10^4$ oscillations in the observable window $10^{-4}~\mrm{Mpc}^{-1}<k<10~\mrm{Mpc}^{-1}$).
These numbers would seem to suggest that in order to measure this signal, measurements which are about 9 orders of magnitude more sensitive than what is currently possible would be needed.
Moreover, it would seem necessary to measure at $\sim10^4$ different $k$ values in order to fully resolve the oscillations.
This seems particularly challenging, given furthermore that fast oscillations run the risk of being washed out in the processes of binning and computing derived power spectra (such as primordial power spectra) from measured CMB power spectra.

Fortunately, not quite as much accuracy should be needed to test the predictions.
In particular, it should be possible to place constraints on the values of $\Omega$ and $\epsilon_\star$ even with existing data, given the scaling of the predicted correction with these parameters.\footnote{In the absence of an independent measurement of the first slow-roll parameter at the pivot scale, $\epsilon_\star$, constraints on $\Omega$ will take the form of a joint constraint on $\Omega$ and $\epsilon_\star$, since existing data currently only provides an upper bound on $\epsilon_\star$.}
As illustrated in Fig.~\ref{fig:Amp_and_Nosc}, lowering the cutoff away from the Planck scale toward the Hubble scale increases the oscillations' amplitude and decreases their frequency, with the effect becoming quite drastic as $\Omega$ approaches the Hubble scale.
Lower values of $\epsilon_\star$ result in smaller-amplitude oscillations, which are unfavourable for detectability, but also slower oscillation frequencies, which can counteract the accelerating effect of having the cutoff approach the Planck scale.
If the oscillations are too slow, however, the correction $\delta \Delta_\mathcal{R}^2$ tends to a constant offset in $\Delta_\mathcal{R}^2$, which is indistinguishable from a change of the Hubble scale $H$ during inflation.
Altogether and at the end of the day, we hope to extract bounds on the possible values of a sharp cutoff $\Omega$ and the slow-roll parameter $\epsilon_\star$ by comparing the predictions to data.

\begin{figure}[t]
    \centering
    \includegraphics[width=0.48\linewidth]{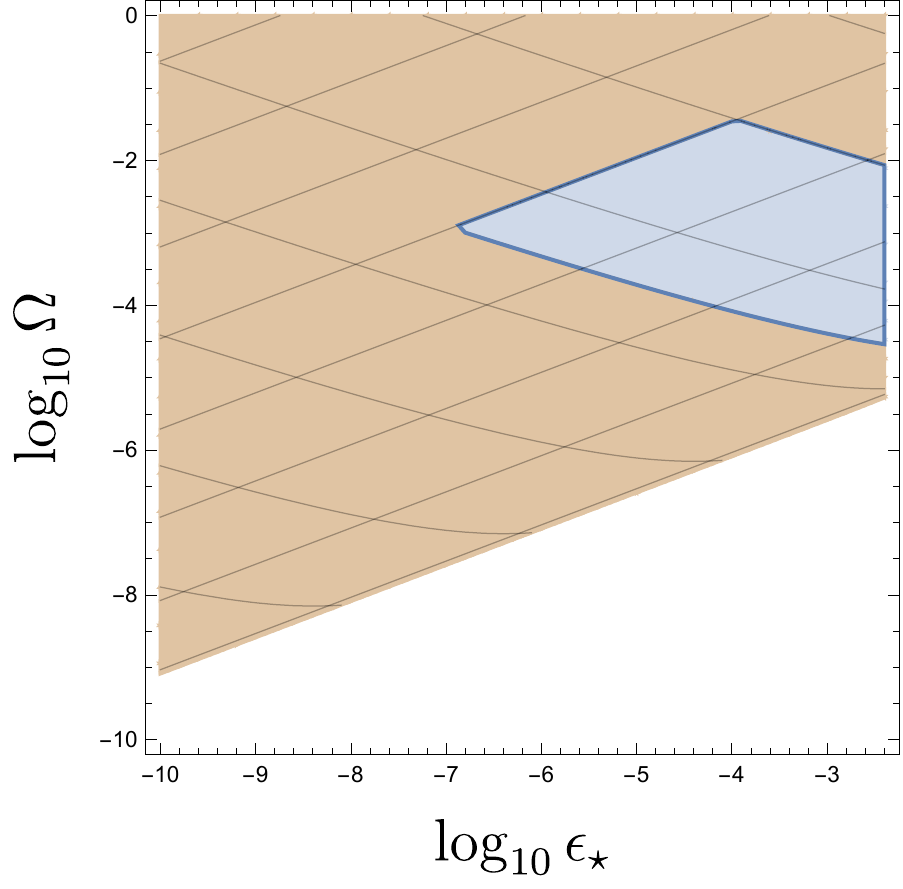}
    \caption{The blue region consists of values of $\epsilon_\star$ and $\Omega$ for which the amplitude of the oscillations satisfies $\log_{10} A > -8$ and for which the number of oscillations in the window of visible modes satisfies $-1 < \log_{10} N_\text{osc} < 2$. The black contour lines match those in Fig.~\ref{fig:Amp_and_Nosc}. We emphasize that these amplitude and frequency thresholds have been chosen arbitrarily and are only meant to illustrate the sorts of bounds that may result from fits to real data.}
    \label{fig:exclusion}
\end{figure}

To get a sense of what these bounds might look like, let us suppose that oscillations whose amplitude is above some threshold and whose frequency lies in some window are detectable.
These ranges of amplitudes and frequencies correspond to specific regions in the plots of \Fig{fig:Amp_and_Nosc}.
Superposing these plots and taking the intersection of the aforementioned regions leads to a contour such as the one shown in \Fig{fig:exclusion}, which illustrates the region of the $\Omega - \epsilon_\star$ parameter space which we could exclude.
We emphasize that the thresholds used to draw an exclusion region here are arbitrary and meant only as illustration; a more careful statistical analysis is of course needed in order to extract bounds, which we leave to future work.
Nevertheless, \Fig{fig:exclusion} illustrates what these bounds might look like.
It is worth mentioning that, in the best possible case in which the predicted oscillations are resolved, this measurement would determine the values of both $\Omega$ and $\epsilon_\star$.

In addition, observational prospects may be aided by the specificity of the prediction.
In particular, given the tightness of the constraints on $A_s$, the frequency of the predicted oscillations in the primordial power spectra depends essentially only on $\Omega$ and $\epsilon_\star$, even if the cutoff is softened. If the cutoff is sharp, then even the phase and amplitude of the predicted oscillations depend only on $\Omega$ and $\epsilon_\star$. This means that the prediction can be thought of as a two-parameter family of template waveforms, parameterized by $\Omega$ (or the ratio of the natural UV cutoff scale to the Hubble scale) and $\epsilon_\star$. 
Experimentally then, the search for these template waveforms should offer a significantly improved signal to noise ratio as, in effect, template search allows one to filter out the noise from that part of the function space that is orthogonal to the space spanned by the template functions, similar to using a low-pass filter to remove noise with frequencies above a signal's frequencies. 
Template search methods have of course recently been used to great effect in the detection of gravitational waves, e.g., from black hole mergers. There, a three-parameter family of template waveforms was successfully used to detect the passage of gravitational waves with a strain of only $10^{-24}$ \cite{LIGOScientific:2016aoc, Privitera:2013xza}. 

It will be very interesting to compare the predictions here with present and upcoming observation-based precision reconstructions of the scalar primordial power spectrum. This will yield at least ever higher lower bounds on the location of a natural UV cutoff scale $\Omega$ and it may eventually yield positive evidence for the existence of a natural UV cutoff, a possible experimental signal of quantum gravitational origin.  

\begin{center} 
{\bf Acknowledgments}
\end{center}
\noindent 
We thank Panos Betzios, Fran\c cois Bouchet, Richard Easther, Simon Foreman, Lukas Hergt, Arjun Kar, Jorma Louko, Rob Martin, and Mark Van Raamsdonk for helpful discussions during the preparation of this manuscript, as well as Gianluca Calcagni and Albert Roura for comments on the first version.
A.C.D. acknowledges the support of the Natural Sciences and Engineering Research Council of Canada (NSERC), [funding reference number PDF-545750-2020].
A.C.D. was supported for a portion of this work as a postdoctoral fellow (Fundamental Research) of the National Research Foundation -- Flanders (FWO), Belgium. AK acknowledges support through a Discovery Grant of the National Science and Engineering Council of Canada (NSERC) and a Discovery Project grant of the Australian Research Council (ARC). PS acknowledges support from the NSERC CGS-D award. 

\appendix

\section{Equivalence between path integrals and projectors}
\label{app:PI-projector-equivalence}

Here, we show that the two expressions for $G_F^\Omega$ given in Eqs.~\eqref{eq:cutoff-GF-PI} and \eqref{eq:GFc} are exactly equivalent for scalar field actions of the form \eqref{eq:scalar-action}.
Starting with \Eq{eq:GF}, act on its left and right with projectors $P_\Omega$:
\begin{align*}
    i (P_\Omega G_F P_\Omega)(x,x') &= i \iint \dee y \, \dee z ~ P_\Omega(x,y) G_F(y,z) P_\Omega(z,x') \\
    &= \frac{1}{\mathcal{N}} \int \mathcal{D}\phi \left( \int \dee y~P_\Omega(x,y) \phi(y) \right) \left( \int \dee z~\phi(z)P_\Omega(z,x')  \right) e^{iS[\phi]}
\end{align*}
In the above, we denote by $\mathcal{N}$ the normalization
\begin{equation}
    \mathcal{N} = \int \mathcal{D}\phi ~ e^{iS[\phi]},
\end{equation}
and we abbreviate the spacetime integral measures, for example $\dee^{d+1}y \sqrt{-g(y)}$ by simply $\dee y$.
Next, because $P_\Omega$ is symmetric, it follows that $P(z,x') = P(x',z)$, and so we may write
\begin{equation} \label{eq:PI-step1}
    i (P_\Omega G_F P_\Omega)(x,x') = \frac{1}{\mathcal{N}} \int \mathcal{D}\phi ~ P_\Omega \phi(x) P_\Omega \phi(x') e^{iS[\phi]} .
\end{equation}
Suppose now that $S[\phi]$ is of the form \eqref{eq:scalar-action}, and insert a resolution of the identity $I = P_\Omega + P_\Omega^\perp$:
\begin{equation} \label{eq:S-res-ID}
\begin{aligned}
    S[(P_\Omega + P_\Omega^\perp)\phi] &= \int \dee x ~ (P_\Omega \phi) F(\Box) (P_\Omega \phi) + \int \dee x ~ (P_\Omega^\perp \phi) F(\Box) (P_\Omega^\perp \phi) \\
    & \qquad + \int \dee x ~ (P_\Omega \phi) F(\Box) (P_\Omega^\perp \phi) + \int \dee x ~ (P_\Omega^\perp \phi) F(\Box) (P_\Omega \phi)
\end{aligned}
\end{equation}
Consider one of the cross terms in the second line of \Eq{eq:S-res-ID}, and expand the integrand in terms of eigenfunctions of the d'Alembertian, $\psi_\lambda$.
Explicitly, writing
\begin{equation}
    \phi(x) = \sum_{\lambda \, \in \, \mrm{spec} \, \Box} \phi_\lambda \psi_\lambda(x),
\end{equation}
we find the following:
\begin{align*}
    \int \dee x ~ (P_\Omega \phi) F(\Box) (P_\Omega^\perp \phi) &= \int \dee x ~ \left( \sum_{|\lambda|\leq \Omega^2} \phi_\lambda \psi_\lambda(x)  \right) F(\Box) \left( \sum_{|\lambda'| > \Omega^2} \phi_{\lambda'} \psi_{\lambda'}(x)  \right) \\[2mm]
    &= \sum_{|\lambda|\leq \Omega^2} \sum_{|\lambda'| > \Omega^2} \phi_\lambda \phi_{\lambda'} F(\lambda') \int \dee x ~ \psi_\lambda(x) \psi_{\lambda'}(x) \\[2mm]
    &= \sum_{|\lambda|\leq \Omega^2} \sum_{|\lambda'| > \Omega^2} \phi_\lambda \phi_{\lambda'} F(\lambda') \delta_{\lambda \lambda'}
\end{align*}
Of course, the sums above are shorthand for a sum over spectrum and should be replaced with integrals wherever $\Box$ has continuous spectrum.
To go to the final line, we used the fact that the d'Alembertian's eigenfunctions are orthonormal.
However, notice that the values of $\lambda$ and $\lambda'$ that are summed over do not overlap, and so the cross-terms in \Eq{eq:S-res-ID} vanish.
We therefore have that $S[\phi] = S[P_\Omega \phi] + S[P_\Omega^\perp \phi]$.

Since $P_\Omega \phi$ and $P_\Omega^\perp \phi$ are independent degrees of freedom, we can path-integrate over them separately.
Therefore, \Eq{eq:PI-step1} altogether reads
\begin{equation}
    i (P_\Omega G_F P_\Omega)(x,x') = \frac{1}{\mathcal{N}} \int \mathcal{D}(P_\Omega \phi) ~ P_\Omega \phi(x) P_\Omega \phi(x') e^{iS[P_\Omega \phi]}  \int \mathcal{D}(P_\Omega^\perp \phi) e^{iS[P_\Omega^\perp \phi]}.
\end{equation}
Similarly, it follows that
\begin{equation}
    \mathcal{N} = \int \mathcal{D}(P_\Omega\phi) ~ e^{iS[P_\Omega \phi]} \int \mathcal{D}(P_\Omega\phi^\perp) ~ e^{iS[P_\Omega^\perp \phi]},
\end{equation}
and so we recover the path integral expression \eqref{eq:cutoff-GF-PI} for the covariantly bandlimited Feynman propagator.

\section{Bandlimited scalar fluctuations in de Sitter}
\label{app:details}

An important part of \Sec{sec:dS-fluctuation-spectrum} was determining the relative change in the fluctuation power spectrum of a massless scalar field, $\hat \phi$, in de Sitter spacetime:
\begin{equation}
\frac{\delta \Delta_\phi^2}{\Delta_\phi^2} = \mrm{Re} \left( \frac{\delta G_F}{G_F} \right).
\end{equation}
$G_F$ denotes the field's Feynman propagator and $\delta G_F = P_\Omega^\perp G_F P_\Omega^\perp - P_\Omega^\perp G_F - G_F P_\Omega^\perp$ is the correction induced by a covariant cutoff, $\Omega$.
We only outlined the six major steps of this calculation in the main text; here we go through each of these steps in full detail.
These steps are:
\begin{enumerate}
    \item Determining the spectrum of the $k$-d'Alembertian, $\Box_k$.
    \item Determining the self-adjoint realizations of $\Box_k$.
    \item Fixing the choice of self-adjoint realization.
    \item Determining the continuous spectrum eigenfunctions.
    \item Using the projector $P_\Omega^\perp$ to compute $\delta \Delta_\phi^2/\Delta_\phi^2$.
    \item Making approximations for calculating $P_\Omega^\perp \chi_0$.
\end{enumerate}

\subsubsection*{Step 1: Determining the spectrum of the $k$-d'Alembertian, $\Box_k$}

Consider the flat slicing of de Sitter spacetime in four spacetime dimensions.
Explicitly, we take the line element to be
\begin{equation}
    \dee s^2 = a^2(\eta)\left[-\dee \eta^2 + \dee x_i \dee x^i \right]
\end{equation}
and the scale factor to be $a(\eta) = (-H\eta)^{-1}$, $\eta \in (-\infty, 0)$.
With these choices, the Sturm-Liouville eigenvalue problem $\Box_k u = \lambda u$ reads
\begin{equation} \label{eq:ee}
    (a^2 u')' + k^2 a^2 u + \lambda a^4 u = 0,
\end{equation}
where $'$ denotes a derivative with respect to the conformal time $\eta$ and where we recall that $k = |{\bm k}|$ is the Fourier variable for $\bm x$.
The eigenfunction equation \eqref{eq:ee} has two linearly independent solutions,
\begin{align}
    u_J(\eta) &= (-\eta)^{3/2} J_{p(\lambda)}(-k\eta) \label{eq:SL-J}\\
    u_Y(\eta) &= (-\eta)^{3/2} Y_{p(\lambda)}(-k\eta), \label{eq:SL-Y}
\end{align}
where $J$ and $Y$ denote Bessel functions of the first and second kind, respectively, and where
\begin{equation} \label{eq:pfunc}
    p(\lambda) = \sqrt{\frac{9}{4}-\frac{\lambda}{H^2}}.
\end{equation}
Because we will ultimately be concerned with a realization of $\Box_k$ as a self-adjoint operator and because the spectra of self-adjoint operators are real, we need only consider real $\lambda$.
Notice that when $\lambda < 9H^2/4$, $u_J$ is normalizable while $u_Y$ is not.
Therefore, self-adjoint realizations of $\Box_k$ will have point spectrum in this range.
Both solutions are non-normalizable for $\lambda \geq 9H^2/4$, and so this range will be the continuous spectrum.

\subsubsection*{Step 2: Determining the self-adjoint realizations of $\Box_k$}

A subtlety that we must address, however, is that the differential expression $\Box_k$ has no unique realization as a self-adjoint operator on $L^2((-\infty,0), a^4(\eta) \, \dee \eta)$.
As a symmetric operator, its deficiency indices are $(1,1)$, meaning that $\Box_k$ has a one-parameter family of self-adjoint extensions corresponding to different (generalized) boundary conditions \cite{naimark1968linear,amrein2005sturm,zettl2012sturm}.
That the deficiency indices are $(1,1)$ can be shown directly by inspecting the solutions of $\Box_k u = \pm i u$, i.e., Eqs.~\eqref{eq:SL-J} and \eqref{eq:SL-Y} with $\lambda = \pm i$.
In both cases, $u_J$ is normalizable while $u_Y$ is not; each deficiency space is therefore one-dimensional and the deficiency indices are $(1,1)$.

A convenient way of parameterizing these self-adjoint extensions is by the location of the largest eigenvalue in the point spectrum.
In particular, a set of square-integrable orthonormal eigenfunctions is given by
\begin{equation} \label{eq:pt_spec_eigenf}
    \psi_n(\eta) = H^2 \sqrt{2 p_n} (-\eta)^{3/2} J_{p_n}(-k\eta),
\end{equation}
where $p_n = p_0 + 2n$, $n \in \mathbb{Z}$, $n \geq 0$, and $p_0 \in (0,2]$.
The tower of discrete eigenvalues in the point spectrum is then
\begin{equation} \label{eq:ptspec}
    \lambda_n = H^2\left(\frac{9}{4} - p_n^2\right),
\end{equation}
and so different choices of $p_0$ produce different point spectra that correspond to the different self-adjoint extensions of $\Box_k$.
Orthonormality of the point spectrum eigenfunctions \eqref{eq:pt_spec_eigenf} follows immediately from computing $\langle \psi_m, \psi_n \rangle$, where the inner product is given by
\begin{equation}
    \langle u, v \rangle = \int_{-\infty}^0 \dee \eta \, a^4(\eta) ~ u^*(\eta) v(\eta)
\end{equation}
for $u, v \in L^2((-\infty,0), a^4(\eta) \dee \eta)$, and from the integral identity \cite[\href{https://dlmf.nist.gov/10.22.E57}{Eq.~10.22.57}]{NIST:DLMF}
\begin{equation}
    \int_0^\infty \frac{\dee x}{x} J_m(x) J_n(x) = \frac{2 \sin(\tfrac{\pi}{2}(m-n))}{\pi(m^2-n^2)} .
\end{equation}

\subsubsection*{Step 3: Fixing the choice of self-adjoint realization}

Having identified the different self-adjoint extensions of $\Box_k$, the obvious question is which one should we use?
We will choose the self-adjoint extension that corresponds to choosing the Bunch-Davies vacuum state for the scalar field, and we will determine the value of $p_0$ to which this corresponds by examining the action of the full (unmodified) Feynman propagator on a test eigenfunction.

Before getting into the above procedure and its meaning, we first need to write down the Feynman propagator.
According to canonical quantization, one can express the Feynman propagator as a time-ordered expectation value,
\begin{equation}
    G_F(x,x') = \bra{0} \mathcal{T} \hat\phi(x) \hat\phi(x') \ket{0}.
\end{equation}
In the conformal coordinates of \Eq{eq:FLRW-line-element}, making a Fourier transform with respect to ${\bm x} - {\bm x}'$, and choosing the Bunch-Davies vacuum state for the mode functions of $\hat \phi$, one arrives at \cite{Birrell:1982ix}
\begin{equation} \label{eq:GF-dS}
\begin{aligned}
    G_F(\eta,\eta';k) = -\frac{i \pi}{4} \frac{\sqrt{\eta\eta'}}{a(\eta)a(\eta')}&\left[\theta(\eta-\eta') H_{3/2}^{(1)}(-k\eta) H_{3/2}^{(2)}(-k\eta') \right. \\ & \left. \qquad + \theta(\eta'-\eta) H_{3/2}^{(2)}(-k\eta) H_{3/2}^{(1)}(-k\eta') \right],
\end{aligned}
\end{equation}
where $H^{(1)}$ and $H^{(2)}$ denote Hankel functions of the first and second kind, respectively.
For reasons that will soon become apparent, note that the propagator above has both Hermitian and anti-Hermitian parts, given by
\begin{equation} \label{eq:GF-herm}
    G_F^h(\eta,\eta';k) = \frac{1}{2} \left[ G_F(\eta,\eta';k) + G_F(\eta',\eta;k)^* \right]
\end{equation}
and
\begin{equation} \label{eq:GF-antiherm}
    G_F^{ah}(\eta,\eta';k) = \frac{1}{2} \left[ G_F(\eta,\eta';k) - G_F(\eta',\eta;k)^* \right],
\end{equation}
respectively.

The Feynman propagator is a right inverse of the d'Alembertian, and so an eigenfunction of the d'Alembertian with eigenvalue $\lambda \neq 0$ is also an eigenfunction of the propagator, but with eigenvalue $\lambda^{-1}$.
Given the expression for $G_F$ in \Eq{eq:GF-dS}, we select a particular self-adjoint extension by dialing the value of $p_0 \in (0,2]$ (which parameterizes the choice of self-adjoint extension) so that the action of $G_F$ on a test eigenfunction of the form \eqref{eq:pt_spec_eigenf} is equal to the same eigenfunction, but multiplied by $\lambda^{-1}$.
In other words, the choice of the Bunch-Davies vacuum in canonical quantization implies a particular choice of self-adjoint extension in the functional analytic language that we are using here, and we are deducing what that choice is.

An important subtlety is that \eqref{eq:GF-dS} is not the integral kernel of an operator on the Hilbert space $L^2((-\infty,0),a^4(\eta) \, \dee \eta)$.
As operators, the d'Alembertian and Feynman propagator satisfy
\begin{equation}
    \hat \Box_k \hat G_{F(k)} = \hat{I},
\end{equation}
where we use a hat to explicitly indicate that this is an operator relation on $L^2((-\infty,0),a^4(\eta) \, \dee \eta)$.
However, the integral kernel $G_F(\eta,\eta';k)$ in \eqref{eq:GF-dS} does not define a good operator on this Hilbert space.
It is straightforward to check that its anti-Hermitian part, \Eq{eq:GF-antiherm}, maps elements of $L^2((-\infty,0),a^4(\eta) \, \dee \eta)$ outside of the Hilbert space.
Rather, it is the Hermitian kernel \eqref{eq:GF-herm} that defines an integral operator whose domain and range are both contained in $L^2((-\infty,0),a^4(\eta) \, \dee \eta)$.

Once a self-adjoint extension has been fixed, each $k$-d'Alembertian can be expressed in terms of its spectrum and eigenfunctions as
\begin{equation}
    \hat \Box_k = \sum_{\mrm{spec}~\Box_k} \lambda~ \ketbra{\lambda}{\lambda},
\end{equation}
where we adopt bra-ket notation to denote vectors and their dual linear functionals.
As its right inverse, we may therefore write
\begin{equation} \label{eq:GFoperator}
    \hat G_{F(k)} = \sum_{\mrm{spec}~\Box_k \setminus \{0\}} \frac{1}{\lambda} ~ \ketbra{\lambda}{\lambda} + (\ketbra{0}{f} + \ketbra{f}{0}).
\end{equation}
If zero is in the spectrum of the chosen self-adjoint extension of the d'Alembertian, then the Feynman propagator's range can have support on this eigenspace as an operator since it is only a right-inverse of the d'Alembertian (i.e. in this case the vector $\ket{f}$ is nonzero).
This is the role played by the bracketed term in \Eq{eq:GFoperator} above.
(Note that the term is written out in a symmetric way, to make manifest the fact that $\hat G_{F(k)}$ is Hermitian.)
We will find that this is indeed the case.

Let $\psi_n(\eta)$ be a test eigenfunction as given by \Eq{eq:pt_spec_eigenf} with $p_0$ not yet fixed.
According to the spectral expansion \eqref{eq:GFoperator}, the action of the (Hermitian part of the) Feynman propagator on $\psi_n(\eta)$ must give
\begin{equation} \label{eq:propagator-condition}
    (G_F^h \psi_n)(\eta) \overset{!}{=} \frac{1}{\lambda_n} \psi_n(\eta) + \alpha (-\eta)^{3/2} J_{3/2}(-k\eta)
\end{equation}
where the constant $\alpha$ may not vanish if $\lambda = 0$ is in the point spectrum.
Let us evaluate the left-hand side:
\begin{align*}
    (G_F^h \psi_n)(\eta) &\equiv \int_{-\infty}^0 a^4(\xi)\, \dee \xi~ G_F^h(\eta,\xi;k) \psi_n(\xi) \\
    &= -\frac{\pi}{4} \sqrt{2 p_n} (-\eta)^{3/2} \left\{ J_{3/2}(-k\eta) \int_0^\infty \frac{\dee x}{x} Y_{3/2}(x) J_{p_n}(x) \right. \\
    &\qquad\qquad\qquad\qquad\qquad - Y_{3/2}(-k\eta) \int_0^\infty \frac{\dee x}{x} J_{3/2}(x) J_{p_n}(x)\\
    &\qquad\qquad\qquad\qquad -2 J_{3/2}(-k\eta) \int_0^{-k\eta} \frac{\dee x}{x} Y_{3/2}(x) J_{p_n}(x) \\
    &\left. \qquad\qquad\qquad\qquad\qquad - 2 Y_{3/2}(-k\eta) \int_0^{-k\eta} \frac{\dee x}{x} J_{3/2}(x) J_{p_n}(x) \right\} 
\end{align*}
In the second equality, we changed the integration variable to $x = -k\xi$.
The following two antiderivatives are useful, assuming that $m^2 \neq n^2$ \cite[\href{https://dlmf.nist.gov/10.22.E6}{Eq.~10.22.6}]{NIST:DLMF}:
\begin{equation}
    \int \frac{\dee x}{x} J_m(x) J_n(x) = \frac{x(J_{m-1}(x) J_n(x) - J_m(x) J_{n-1}(x)) - (m-n)J_m(x)J_n(x)}{m^2-n^2}
\end{equation}
\begin{equation}
    \int \frac{\dee x}{x} Y_m(x) J_n(x) = \frac{x(Y_{m-1}(x) J_n(x) - Y_m(x) J_{n-1}(x)) - (m-n)Y_m(x)J_n(x)}{m^2-n^2}
\end{equation}
In particular, both antiderivatives vanish as $x \rightarrow 0^+$ if $m < n$.
Therefore, let us choose $p_n$ with $n \geq 1$ so that the $x=0$ endpoint of the integrals above do not contribute.
Making this choice, we arrive at
\begin{equation}
\begin{aligned}
    (G_F^h \psi_n)(\eta) &= \frac{1}{\lambda_n} \psi_n(\eta) + \frac{1}{2(\tfrac{9}{4}-p_n^2)} \sqrt{2 p_n} (-\eta)^{3/2} \\
    & \qquad \cdot \left\{ -J_{3/2}(-k\eta) \cos(\tfrac{\pi}{2}(\tfrac{3}{2}-p_n)) + Y_{3/2}(-k\eta)  \sin(\tfrac{\pi}{2}(\tfrac{3}{2}-p_n)) \right\}.
\end{aligned}
\end{equation}
The first term in the equation above is what we expect from inverse action, and the second term is in the kernel of the d'Alembertian \emph{as a differential expression}.
However, the $Y_{3/2}$ contribution is not normalizable, and so it must vanish if we require the Hermitian part of the Feynman propagator \emph{as an operator} to map into the Hilbert space.
This happens when $p_0 = 3/2$, giving $\sin(\tfrac{\pi}{2}(\tfrac{3}{2}-p_n)) = 0$.
Altogether, we then arrive at
\begin{equation}
    (G_F^h \psi_n)(\eta) = \frac{1}{\lambda_n} \psi_n(\eta) - \frac{(-1)^n}{\lambda_n} \sqrt{\frac{p_n}{6}}  \psi_0(\eta),
\end{equation}
and we have concluded that the self-adjoint extension implicit in the choice of Bunch-Davies vacuum for the Feynman propagator is the one for which $p_0 = 3/2$.
Notice that $\lambda_0 = 0$ is therefore in the point spectrum.

\subsubsection*{Step 4: Determining the continuous spectrum eigenfunctions}

Once we have fixed the self-adjoint extension of $\Box_k$ as well as the orthonormal eigenvectors in the point spectrum, we can determine the eigenfunctions for $\lambda > 9H^2/4$ in the continuous spectrum.\footnote{We will not need the edge case $\lambda = 9H^2/4$, for which $\varphi_0(\eta)$ can be expressed as a linear combination of $J_0(-k\eta)$ and $Y_0(-k\eta)$.}
Let us take as an ansatz
\begin{equation}
    \varphi_q(\eta) = H^2 (-\eta)^{3/2} N_q (a_q \, \Re \, J_{iq}(-k\eta) + b_q \, \Im \, J_{iq}(-k\eta)),
\end{equation}
where
\begin{equation} \label{eq:qfunc}
    q \equiv q(\lambda) = \sqrt{\frac{\lambda}{H^2}-\frac{9}{4}},
\end{equation}
and the weights $a_q$, $b_q$ as well as the overall normalization $N_q$ are all real.
$J_{iq}$ is a Bessel function of the first kind of purely imaginary order, and we use its real and imaginary parts to form linearly independent solutions of the eigenfunction equation.

First, we can determine $a_q$ and $b_q$ by demanding that the $\psi_n$ and $\varphi_q$ must be orthogonal:
\begin{align*}
    \langle \psi_n, \varphi_q \rangle &= \int_{-\infty}^0 a^4(\eta) \, \dee \eta ~ H^4 (-\eta)^3 \sqrt{2 p_n} N_q J_{p_n}(-k\eta) (a_q \, \Re \, J_{iq}(-k\eta) + b_q \, \Im \, J_{iq}(-k\eta)) \\[2mm]
    &\propto \int_0^\infty \frac{\dee x}{x} J_{p_n}(x) (a_q \, \Re \, J_{iq}(x) + b_q \, \Im \, J_{iq}(x))
\end{align*}
In the second line we again defined $x = -k\eta$.
According to \cite[\href{https://dlmf.nist.gov/10.22.E57}{Eq.~10.22.57}]{NIST:DLMF}, we have that
\begin{equation}
\begin{aligned}
    \int_0^\infty \frac{\dee x}{x} J_{p_n}(x) J_{iq}(x) &= \frac{\Gamma(\tfrac{p_n}{2}+\tfrac{iq}{2})}{2\Gamma(1+\tfrac{p_n}{2}+\tfrac{iq}{2})\Gamma(1+\tfrac{p_n}{2}-\tfrac{iq}{2})\Gamma(1-\tfrac{p_n}{2}+\tfrac{iq}{2})} \\[2mm]
    &= \frac{\sin\left(\tfrac{\pi}{2}(p_n - iq)\right)}{\tfrac{\pi}{2}(p_n^2+q^2)} \\[2mm]
    &= \frac{(-1)^n \sqrt{2}}{\pi(p_n^2+q^2)} \left( \cosh(\tfrac{\pi}{2}q) + i \sinh(\tfrac{\pi}{2}q) \right).
\end{aligned}
\end{equation}
To go to the second line, we used gamma function identities, and in the third line we used that $p_n = \tfrac{3}{2} + 2n$.
Consequently,
\begin{equation}
    \langle \psi_n, \varphi_q \rangle \propto a_q \cosh(\tfrac{\pi}{2}q) + b_q \sinh(\tfrac{\pi}{2}q).
\end{equation}
Motivated by hindsight, a convenient choice is $a_q = \sech(\tfrac{\pi}{2} q)$ and $b_q = -\csch(\tfrac{\pi}{2} q)$, giving
\begin{equation}
    \varphi_q(\eta) = H^2 (-\eta)^{3/2} N_q \left[ \sech(\tfrac{\pi}{2} q) \, \Re \, J_{iq}(-k\eta) - \csch(\tfrac{\pi}{2} q) \, \Im \, J_{iq}(-k\eta) \right].
\end{equation}

To fix the normalization, we will require that
\begin{equation}
    \langle \varphi_q, \varphi_{q'} \rangle = \delta(q - q').
\end{equation}
This amounts to setting $N_q = \sqrt{\tfrac{1}{2} q \tanh(\pi q)}$, which can be verified by numerically integrating $\langle \varphi_q, \varphi_{q'} \rangle$ around a small neighbourhood of $q - q' = 0$ or via the following analytic argument.\footnote{We thank Jorma Louko for pointing this out to us.}
On grounds of orthonormality, we know that $\langle \varphi_q, \varphi_{q'} \rangle$ is proportional to $\delta(q-q')$.
As such, when evaluating $\langle \varphi_q, \varphi_{q'} \rangle$, we need only look for contributions to its distributional part.
By definition, we have that
\begin{equation} \label{eq:distrib1}
    \langle \varphi_q, \varphi_{q'} \rangle = \int_0^\infty \frac{\dee x}{x} \, \frac{1}{2} \left[ \sqrt{q \tanh(\pi q)} \left( \sech(\tfrac{\pi}{2} q) \, \Re \, J_{iq}(x) - \csch(\tfrac{\pi}{2} q) \, \Im \, J_{iq}(x)  \right) \right] \cdot \left[ q \rightarrow q' \right].
\end{equation}
For large $x$, $|J_{iq}(x)|/x \leq C/x^{3/2}$ for some constant $C$, and so the integration over any interval $(\epsilon, \infty)$ with $\epsilon > 0$ will not give a distributional contribution.
For small $x$,
\begin{equation}
    J_{iq}(x) = \left(\frac{x}{2}\right)^{iq} \frac{1}{\Gamma(1+iq)} (1 + O(x^2) ).
\end{equation}
Therefore, we may replace $J_{iq}(x)$ with $(x/2)^{iq}/\Gamma(1+iq)$ to compute the distributional contribution coming from the $x=0$ endpoint of integration.
Making this replacement, we find that
\begin{equation} \label{eq:distrib2}
\begin{aligned}
    \sech(\tfrac{\pi}{2} q) \, \Re \, J_{iq}(x) -  \csch(\tfrac{\pi}{2} q) & \, \Im \, J_{iq}(x) \\ & \rightarrow ~ \frac{1}{2} \left[ (\sech(\tfrac{\pi}{2}q) + i \csch(\tfrac{\pi}{2}q)) \left( \frac{x}{2} \right)^{iq} \frac{1}{\Gamma(1+iq)} \right. \\
    & \left. \qquad + (\sech(\tfrac{\pi}{2}q) - i \csch(\tfrac{\pi}{2}q)) \left( \frac{x}{2} \right)^{-iq} \frac{1}{\Gamma(1-iq)} \right].
\end{aligned}
\end{equation}
The distributional part will therefore come from terms of the form
\begin{equation} \label{eq:distrib3}
    \int_0^\epsilon \frac{\dee x}{x} \left( \frac{x}{2} \right)^{ i \gamma} = \int_{\tilde \epsilon}^\infty \dee t ~ e^{-i \gamma t} = \pi \delta(\gamma) + (\text{non-distributional}),
\end{equation}
where we let $x = 2 e^{-t}$ in the first equality.
We can then read off the coefficient of $\delta(q-q')$ from Eqs.~\eqref{eq:distrib1}, \eqref{eq:distrib2}, and \eqref{eq:distrib3}; it is obtained by collecting the coefficients of the cross-terms $(x/2)^{\pm i(q-q')}$, setting $q = q'$, and multiplying by $\pi$, as we are instructed to do by \eqref{eq:distrib3}:
\begin{equation}
\begin{aligned}
    &\pi \cdot \frac{1}{2} q \tanh(\pi q) \cdot \frac{1}{4} \left( 2 \frac{(\sech(\tfrac{\pi}{2}q) + i \csch(\tfrac{\pi}{2}q))(\sech(\tfrac{\pi}{2}q) - i \csch(\tfrac{\pi}{2}q))}{\Gamma(1+iq)\Gamma(1-iq)} \right) \\[2mm]
    &= \frac{\pi}{4} q \tanh(\pi q) \left( \sech^2(\tfrac{\pi}{2}q) + \csch^2(\tfrac{\pi}{2}q) \right) \frac{\sinh(\pi q)}{\pi q} \\[2mm]
    &= 1
\end{aligned}
\end{equation}
In going to the second and third lines, we used standard gamma function and hyperbolic identities, respectively.
Therefore, it follows that
\begin{equation} \label{eq:cts_spec_eigenf}
    \varphi_q(\eta) = H^2 (-\eta)^{3/2} \sqrt{\tfrac{1}{2}q \tanh(\pi q)} \left[ \sech(\tfrac{\pi}{2} q) \, \Re \, J_{iq}(-k\eta) - \csch(\tfrac{\pi}{2} q) \, \Im \, J_{iq}(-k\eta) \right]
\end{equation}
are indeed continuum-orthonormalized in the index $q$.

\subsubsection*{Step 5: Using the projector $P_\Omega^\perp$ to compute $\delta \Delta_\phi^2/\Delta_\phi^2$}

With all of the ingredients in hand, we can now construct the projector $P_\Omega^\perp$.
The (integral kernel of the) projector is given by
\begin{equation} \label{eq:P-Omega-perp}
    P_\Omega^\perp(\eta,\eta') = \int_Q^\infty \dee q ~ \varphi_q(\eta) \varphi_q(\eta') + \sum_{n \geq N} \psi_n(\eta) \psi_n(\eta'),
\end{equation}
where $Q = q(\Omega^2)$, as given by \Eq{eq:qfunc}, and $N = \min \{n : \lambda_n < -\Omega^2\}$ (cf. \Eq{eq:ptspec}).

We must now compute $\delta \Delta^2_\phi / \Delta^2_\phi$, as given by \Eq{eq:reldiffscalar}.
First, note that this expression simplifies if we carefully examine the structure of the Feynman propagator.
The Feynman propagator evaluated at equal times is a purely imaginary quantity,
\begin{equation}
G_F(\eta=\eta';k) = - \frac{i \pi}{4}  (-\eta)^3 H^2 \left( J_{3/2}(-k\eta)^2 + Y_{3/2}(-k\eta)^2 \right).
\end{equation}
Moreover, the Hermitian (\Eq{eq:GF-herm}) and anti-Hermitian (\Eq{eq:GF-antiherm}) parts of the Feynman propagator, and hence also the two correction terms $\delta G_F^h$ and $\delta G_F^{ah}$, are purely real and purely imaginary, respectively.\footnote{In fact, they are purely real and purely imaginary not just at equal times, but for all times.}
Therefore, we have that
\begin{equation} \label{eq:reldiff}
\frac{\delta \Delta^2_\phi}{\Delta^2_\phi} \approx \left. \frac{\delta G_F^{ah}}{G_F} \right|_{\eta=\eta'},
\end{equation}
and so we only need to compute the correction to the anti-Hermitian part of the propagator.

In analogy with the eigenfunction for the eigenvalue $\lambda = 0$, let us define the following function:
\begin{equation} \label{eq:chi0}
\chi_0(\eta) = H^2 \sqrt{3} (-\eta)^{3/2} Y_{3/2}(-k\eta)
\end{equation}
Then, we can write $G_F^{ah}(\eta,\eta';k)$ as
\begin{equation}
    G_F^{ah}(\eta,\eta';k) = -\frac{i \pi}{12 H^2} \left[\psi_0(\eta)\psi_0(\eta') + \chi_0(\eta) \chi_0(\eta')  \right],
\end{equation}
as well as its bandlimited version as
\begin{equation}
\begin{aligned}
    G_F^{ah,\Omega} &= P_\Omega G_F^{ah} P_\Omega \\
    &= G_F^{ah} + (P_\Omega^\perp G_F^{ah} P_\Omega^\perp - P_\Omega^\perp G_F^{ah} - G_F^{ah} P_\Omega^\perp) \\
    &\equiv G_F^{ah} + \delta G_F^{ah}.
\end{aligned}
\end{equation}
Since $\ip{\psi_\lambda}{\psi_0} = 0$ for $\lambda \neq 0$, it is straightforward to show that
\begin{equation}
\delta G_F^{ah}(\eta=\eta') = \frac{i\pi}{6 H^2} (P_\Omega^\perp \chi_0)(\eta) \left[ \chi_0(\eta) - \frac{1}{2}(P_\Omega^\perp \chi_0)(\eta)  \right], \label{eq:dGFah}
\end{equation}
where we explicitly have that
\begin{equation} \label{eq:P-perp-chi}
    (P_\Omega^\perp \chi_0)(\eta) = \int_Q^\infty \dee q ~ \langle \varphi_q, \chi_0 \rangle \varphi_q(\eta) + \sum_{n \geq N} \langle \psi_{n}, \chi_0 \rangle \psi_{n}(\eta).
\end{equation}

\subsubsection*{Step 6: Making approximations for calculating $P_\Omega^\perp \chi_0$}

Next we turn to computing $(P_\Omega^\perp \chi_0)(\eta)$.
Consider first the contribution from the continuous spectrum, i.e., the first term on the right side of \Eq{eq:P-perp-chi}.
The inner product appearing in the integral is given by
\begin{equation}
    \langle \varphi_q, \chi_0 \rangle = \frac{-2\sqrt{3 q \tanh(\pi q)}}{\pi (\tfrac{9}{4} + q^2)},
\end{equation}
and so the integrand reads
\begin{equation} \label{eq:damnintegrand}
    \langle \varphi_q, \chi_0 \rangle \psi_q(\eta) = -\frac{\sqrt{6}}{\pi} \frac{q \tanh(\pi q)}{(\tfrac{9}{4} + q^2)}  H^2 (-\eta)^{3/2} \left[ \sech (\tfrac{\pi}{2} q) \Re J_{iq}(-k\eta) - \csch(
    \tfrac{\pi}{2} q) \Im J_{iq}(-k\eta) \right].
\end{equation}
Unfortunately, this cannot be integrated in closed form, and furthermore, it is totally intractable numerically due to its oscillatory nature.
Fortunately, for large values of $q$, we can make several approximations to make this integrand more tractable.
We approximate
\begin{equation}
    \frac{q \tanh(\pi q)}{\tfrac{9}{4} + q^2} \approx \frac{1}{q},
\end{equation}
as well as
\begin{equation}
    \sech(\tfrac{\pi}{2} q) \approx \csch(\tfrac{\pi}{2} q) \approx 2 e^{-\tfrac{\pi}{2} q}.
\end{equation}
Then,
\begin{equation}
    \langle \psi_q, \chi_0 \rangle \psi_q(\eta) \approx - \frac{2\sqrt{6}}{\pi} H^2 (-\eta)^{3/2} \frac{1}{q} e^{-\tfrac{\pi}{2} q} \left[ \Re J_{iq}(-k\eta) - \Im J_{iq}(-k\eta) \right].
\end{equation}
To make further progress, we can invoke an asymptotic expansion for Bessel functions of large order:
\begin{equation} \label{eq:large-order}
    J_\mu (z) \approx \frac{1}{\sqrt{2 \pi \mu}} \left( \frac{ez}{2\mu}  \right)^{\mu}
\end{equation}
According to \cite[\href{https://dlmf.nist.gov/10.19.E1}{Eq.~10.19.1}]{NIST:DLMF} this expansion holds for large positive orders $\mu$.
However, by inspection, it seems to continue to hold if we analytically continue $\mu = i \nu$.

\begin{figure}[t]
    \centering
    \includegraphics[width=0.45\textwidth]{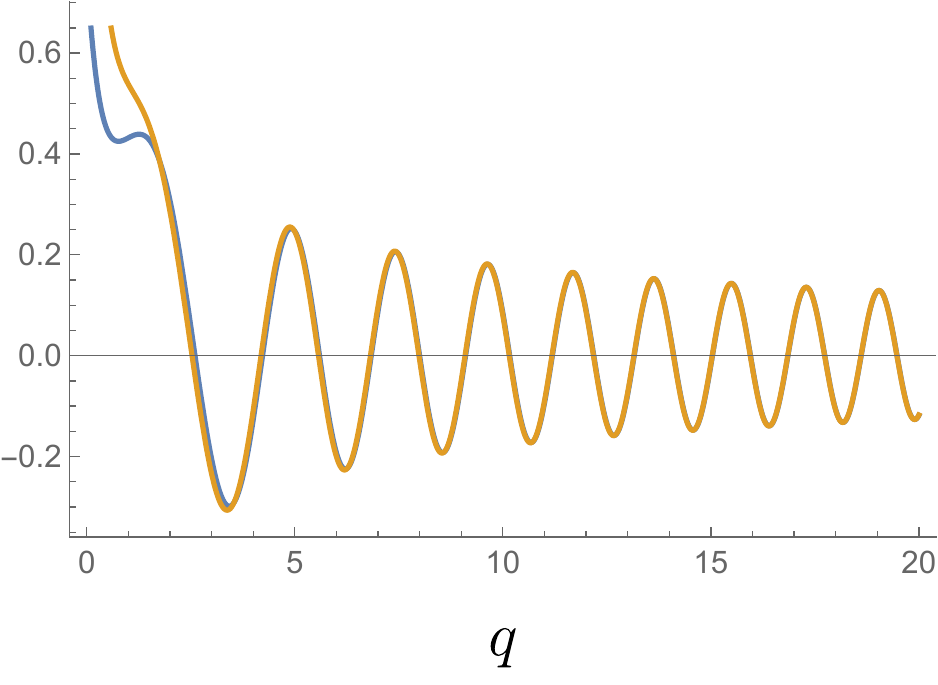}~
    \includegraphics[width=0.45\textwidth]{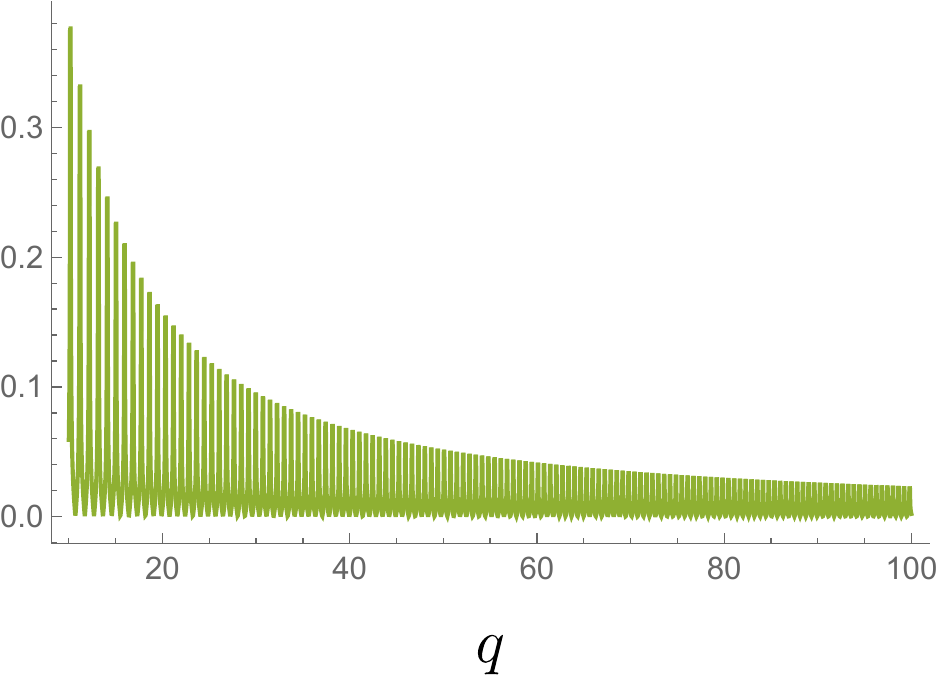}
   \caption{Left: Comparison of $F_1(q) = e^{-(\pi/2) q} (\Re J_{iq}(1) - \Im J_{iq}(1))$ (blue) and $F_2(q) = (\pi q)^{-1/2} \cos (q \ln (e/2q))$ (orange). Right: The relative difference, $|F_1(q)-F_2(q)|/|F_1(q)|$. Beyond $q \sim O(10^2)$, $F_1(q)$ becomes numerically unstable.}
    \label{fig:large-order}
\end{figure}

After a bit of algebraic manipulation, we find that
\begin{equation}
    \Re J_{iq}(z) - \Im J_{iq}(z) \approx \frac{1}{\sqrt{\pi q}} \Re \left[ \left( \frac{ez}{2 i q} \right)^{iq} \right].
\end{equation}
We further have that
\begin{equation}
\begin{aligned}
    \left( \frac{ez}{2 i q} \right)^{iq} &= \exp \left\{ iq \ln\left( \frac{ez}{2iq}  \right) \right\} \\
    &= \exp\left\{ \frac{\pi}{2}q + i q \ln\left( \frac{ez}{2q} \right) \right\},
\end{aligned}
\end{equation}
where we made a branch cut for the complex logarithm.
It therefore follows that
\begin{equation} \label{eq:large-order-applied}
    e^{-\tfrac{\pi}{2} q} (\Re J_{iq}(z) - \Im J_{iq}(z)) \approx \frac{1}{\sqrt{\pi q}} \cos \left( q \ln\left( \frac{ez}{2q} \right)  \right).
\end{equation}
To get a sense of the goodness of the approximation, both sides of \Eq{eq:large-order-applied}, as well as their difference, is plotted in \Fig{fig:large-order}.

Altogether, we arrive at a nice approximate but compact expression for the integrand \eqref{eq:damnintegrand}:
\begin{equation} \label{eq:integrand-approximation}
    \langle \psi_q, \chi_0 \rangle \psi_q(\eta) \approx - \sqrt{3} H^2 \left( - \frac{2 \eta}{\pi q}  \right)^{3/2} \cos \left( q \ln\left( \frac{2q}{-k \eta e} \right)  \right)
\end{equation}
This is a vast improvement, but there is still no closed form expression for the antiderivative
\begin{equation}\label{eq:oscillatory-integral}
    \int \dee q ~ q^{-3/2} \cos(q \ln(A q)) .
\end{equation}
Nevertheless, it converges when integrated over the interval $[Q, \infty)$, due to the overall power of $q^{-3/2}$.
A numerically efficient way of evaluating this integral is to instead perform an equivalent integration along a contour in the complex plane.

\begin{figure}[t]
    \centering
    \includegraphics[width=0.5\textwidth]{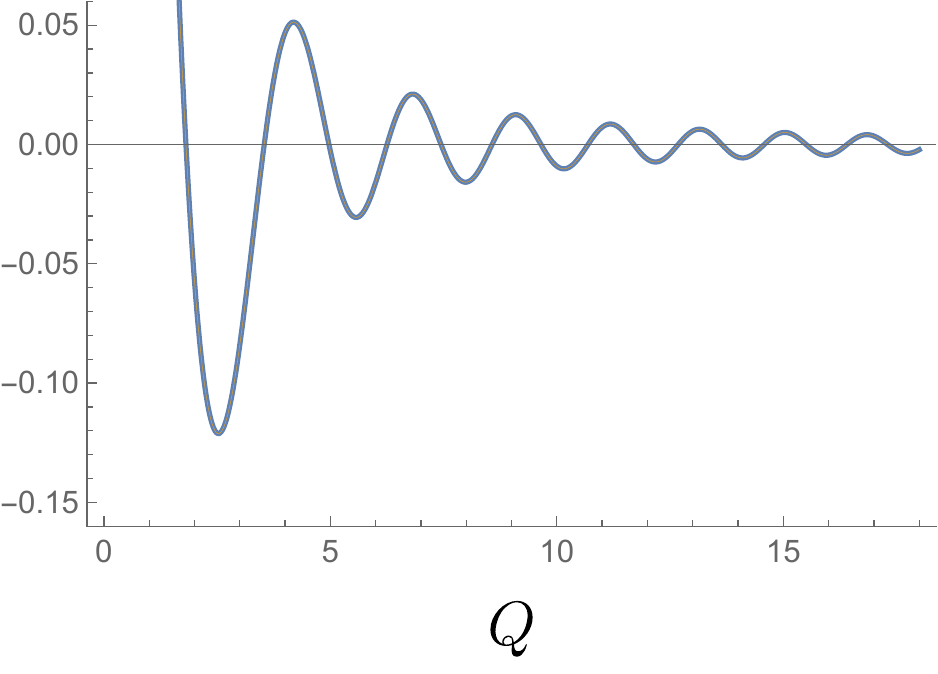}
    \caption{$I(Q,-k\eta)$ and $\int_Q^\infty \dee q ~ q^{-3/2} \cos(q \ln[2q/(-k\eta e)])$ are virtually indistinguishable, plotted here with $-k\eta = 1$ and for a range of $Q$ over which it is still possible to numerically integrate the latter. Their relative difference is not shown, as it is at the level of machine precision.}
    \label{fig:IQA}
\end{figure}

Define the integral
\begin{equation}
    \tilde I(Q,A) = \int_Q^\infty  \dee q ~ e^{i q \ln(A q) - \tfrac{3}{2} \ln q} ,
\end{equation}
so that its real part is the integral that we wish to compute.
Denote the argument of the exponential by
\begin{equation}
    \mathcal{I}(q,A) = i q \ln(A q) - \frac{3}{2} \ln q .
\end{equation}
Its only singularities are at $q = 0$ and $q \rightarrow \infty$, and so taking the branch cut of the complex logarithm along the negative real axis, we can safely deform the domain of integration to a contour which starts at $q = Q$ and goes up vertically into the complex plane, to wit,
\begin{equation}
    \tilde I(Q,A) = \int_{q = Q}^{q = Q + i \infty} \dee q ~ e^{\mathcal{I}(q,A)} = \int_0^\infty i \, \dee b ~ e^{\mathcal{I}(Q + ib,A)} .
\end{equation}
Write $q = a + ib$ and split $\mathcal{I}$ into its real and imaginary parts by writing
\begin{equation}
    \mathcal{I}(a + ib,A) = w(a,b;A) + i W(a,b;A) .
\end{equation}
One then finds that
\begin{align}
    w(a,b;A) &= - \frac{1}{2} \left(b + \frac{3}{2}\right) \ln\left(a^2+b^2\right) - a \arctan\left(\frac{b}{a}\right) - b \ln A \label{eq:little-w}\\
    W(a,b;A) &= \frac{a}{2} \ln\left(a^2+b^2\right) - \left(b + \frac{3}{2}\right) \arctan\left(\frac{b}{a}\right) + a \ln A . \label{eq:big-W}
\end{align}
We only need the real part of $\tilde I$; therefore, we need only compute
\begin{equation} \label{eq:I-integral-app}
\begin{aligned}
    \Re \left[\tilde I(Q,2/(-k\eta e))\right] &= - \int_0^\infty \dee b ~ e^{w(Q,b;2/(-k\eta e))} \sin\left( W[Q,b;2/(-k\eta e)] \right) \\[2mm]
    &\equiv I(Q,-k\eta) .
\end{aligned}
\end{equation}
In particular, notice that even though there is an oscillatory component to this integral, the $h$-dependent prefactor decays exponentially quickly with $y$, and so this integral converges extremely rapidly.
$I(Q,-k\eta)$ is virtually indistinguishable from $\int_Q^\infty \dee q ~ q^{-3/2} \cos(q \ln[2q/(-k\eta e)])$, as shown in \Fig{fig:IQA}, and the difference between the two is at the level of machine precision.

Altogether, the projection $(P_\Omega^\perp \chi_0)(\eta)$ therefore contains a contribution from the continuous spectrum given by
\begin{equation} \label{eq:Pperp}
    (P_\Omega^\perp \chi_0)(\eta) \supset - \sqrt{3} H^2 \left( - \frac{2 \eta}{\pi}  \right)^{3/2} I\left(Q,2/(-k\eta e)\right).
\end{equation}

Next we consider the contribution from the point spectrum, i.e., the second term on the right side of \Eq{eq:P-perp-chi}.
Here, the inner product appearing in the summand is given by
\begin{equation}
    \ip{\psi_n}{\chi_0} = - \sqrt{6 p_n} \frac{(-1)^n}{\pi n (3 + 2n)},
\end{equation}
and so $(P_\Omega^\perp \chi_0)(\eta)$ contains a contribution from the point spectrum given by
\begin{equation}
    (P_\Omega^\perp \chi_0)(\eta) \supset - \sqrt{3} H^2 \left( - \frac{2 \eta}{\pi}  \right)^{3/2}  \sum_{n \geq N} \sqrt{\frac{\pi}{2}} \frac{(-1)^n}{n}  \frac{(\tfrac{3}{2}+2n)}{(3+2n)} J_{p_n}(-k\eta).
\end{equation}
Because the sum is an alternating series, its magnitude is bounded from above by
\begin{equation}
    \sqrt{\frac{\pi}{2}} \frac{1}{N}  \frac{(\tfrac{3}{2}+2N)}{(3+2N)} |J_{p_N}(-k\eta)| \approx \frac{1}{(2 N)^{3/2}} \left( \frac{-k\eta e}{4 N} \right)^{2 N},
\end{equation}
where the approximation holds for large $N$.
In particular, we approximated $p_N \approx 2N$ and used the large-order approximation \eqref{eq:large-order}.

This bound is laughably tiny.
For illustration, when $\Omega \gg H$, it follows that $N \approx \Omega/(2H)$, and even for, e.g., $\Omega/H = 100$, the bound evaluates to $7 \times 10^{-194}$ at horizon crossing (when $-k\eta = 1$).
The value of $I(Q,2/e)$ is roughtly $1.1 \times 10^{-4}$, and so we may safely neglect the contribution from the point spectrum.

It is then a matter of straightforward algebraic manipulations to arrive at the expression \eqref{eq:final_answer} for $\delta \Delta_\phi^2/\Delta_\phi^2$.

\section{Adiabatic approximation}
\label{sec:approximation}

Here we elaborate and further justify our use of the adiabatic, or ``slow-roll'' approximation for nearly de Sitter spacetimes. 

Suppose we wish to compute an inflationary observable $\mathcal O_k$ which depends on the FLRW scale factor $a(\eta)$ and the time $\eta_k$ at which the mode $k$ crosses the horizon. That is, $\eta_k$ is defined as the solution of
\begin{align}
    k = a(\eta_k)H(\eta_k),
\end{align}
where $H = a'/a^2$. Hence, we can write $\mathcal O_k = \mathcal O_k[a,\eta_k]$. The adiabatic approximation is the statement that
\begin{align}
    \mathcal O_k[a,\eta_k] \approx \mathcal O_k[\tilde a,\tilde \eta_k],
\end{align}
where $\tilde a(\eta)\equiv -1/\tilde H \eta$ is a de Sitter scale factor. We define the de Sitter Hubble constant $\tilde H$ and the time $\tilde \eta_k$ at which the mode $k$ crosses the de Sitter horizon via the equations
\begin{align}
    k &= \tilde a (\tilde \eta_k)\tilde H,\label{eq:dS_k_crossing}\\
    \tilde a(\tilde \eta_k) &= a(\eta_k).\label{eq:a=adS}
\end{align}
Equation \eqref{eq:dS_k_crossing} is the usual horizon crossing condition for the mode $k$, while \eqref{eq:a=adS} ensures that the de Sitter scale factor is equal to the ``true'' scale factor at the mode crossing time. It is a generic feature of inflationary cosmology that observables associated with a mode $k$ only depend on the FLRW evolution near the time at which the mode $k$ crosses the horizon. Our adiabatic approximation is simply making use of this fact to approximate an observable in a generic FLRW spacetime with the same observable in a de Sitter spacetime, but with the de Sitter spacetime tuned in such a way that near the $k$-mode crossing time it looks similar to the true spacetime. The utility of the approximation arises when it is difficult to compute the observable in the true spacetime, but it is relatively easy to do so in the de Sitter spacetime. 

For example, in this paper we were interested in the case where the observable $\mathcal O_k$ is the covariantly bandlimited scalar or tensor power spectrum. This quantity is unfeasible to compute for a generic FLRW spacetime, but we are able to compute it in the de Sitter case. The adiabatic approximation is therefore useful.

Let us now consider a simpler observable: the scalar power spectrum \textit{without a cutoff.} This is the quantity to which we are computing Planck-scale corrections in this paper. For a FLRW spacetime with scale factor $a(\eta)$ and for a massless scalar field $\phi$, it is defined as 
\begin{align}
    \Delta^2_\phi(k) = \frac{k^3}{2\pi^2 a^2(\eta_k)}|v_k(\eta_k)|^2,
\end{align}
where $v_k$ is the solution to the mode equation
\begin{align}\label{eq:mode_equation}
    v_k''+\left(k^2 - \frac{a''}{a}\right)v_k=0,
\end{align}
with initial condition $v_k\rightarrow \frac{1}{\sqrt{2k}}e^{-i k \eta}$ as $\eta\rightarrow -\infty$ for the Bunch-Davies vacuum. This quantity can be computed exactly for both power law and de Sitter spacetimes, allowing us to use it to explicitly test the validity of the adiabatic approximation. 

In other words, suppose that the ``true'' spacetime is a power law spacetime, with a scale factor
\begin{align}
    a = A t^c = A[A(c-1)(-\eta)]^{c/(1-c)},
\end{align}
where we take $c>2$ so that the spacetime is inflating.
In terms of cosmic time $t$, conformal time is given by \begin{align}
    \eta = -\frac{1}{A(c-1)t^{c-1}}.
\end{align}
The Hubble parameter is
\begin{align}
    H = \frac{c}{t} = c[A(c-1)(-\eta)]^{1/(c-1)}.
\end{align}
Notice that the first slow-roll parameter
\begin{align}\label{eq:epsilon_power_law}
    \epsilon \equiv \frac{d}{dt}\left(\frac{1}{H}\right)=\frac{1}{c}
\end{align}
vanishes in the limit $c\rightarrow\infty$. Thus, we might expect that in this limit the power law spacetime behaves similarly to a de Sitter spacetime. However, even at finite $c$, we will find that the adiabatic approximation can accurately estimate the power spectrum $\Delta_\phi^2$ for a scalar field in the power law spacetime. 

To see this, let us first compute the exact answer for $\Delta_\phi^2$. Taking the power law scale factor, the solution to the mode equation \eqref{eq:mode_equation} with the required initial conditions is
\begin{align}
    v_k(\eta) = -\frac{\sqrt{\pi}}{2}\sqrt{-\eta}H^{(1)}_{\nu_c}(-k\eta),
\end{align}
where $\nu_c = \frac{3}{2}+\frac{1}{c-1}$. Hence, the exact power spectrum is
\begin{align}\label{eq:exact}
    \Delta_\phi^2(k) =
    \frac{c^2}{8\pi} \left(\frac{Ac}{k}\right)^{\frac{2}{c-1}}
    \frac{c}{c-1}
    \left|H^{(1)}_{\nu_c}\left(\frac{c}{c-1}\right)\right|^2.
\end{align}

Let us now approximate the power spectrum using the adiabatic approximation. To do so, we need the expression for $\Delta^2_\phi$ for a de Sitter spacetime. In this case, the mode function solution of \eqref{eq:mode_equation} can be written in closed form as
\begin{align}
    \tilde v_k(\tilde \eta) = -\frac{\sqrt{\pi}}{2}\sqrt{-\tilde\eta}H^{(1)}_{\nu_\infty}(-k\tilde \eta),
\end{align}
where $\nu_\infty = 3/2$. The adiabatic approximation instructs us to set the conformal time to $\eta_k$ and take the de Sitter Hubble constant as $\tilde H$, where $\eta_k$ and $\tilde H$ are solutions to equations \eqref{eq:dS_k_crossing} and \eqref{eq:a=adS}. The solutions to these equations read
\begin{align}
    \eta_k &= \frac{c}{(1-c)k},\\
    \tilde H &= c\left(\frac{Ac}{k}\right)^{\frac{1}{c-1}},
\end{align}
where we have also made use of the fact that the de Sitter horizon crossing time is $\tilde \eta_k = -1/k$.
The adiabatic approximation for the power spectrum is thus
\begin{align}\label{eq:approx}
    \tilde\Delta_\phi^2(k) =
    \frac{c^2}{8\pi} \left(\frac{Ac}{k}\right)^{\frac{2}{c-1}}
    \left|H^{(1)}_{\nu_\infty}(1)\right|^2.
\end{align}

Let us compare the approximate expression $\tilde \Delta_\phi^2$ in equation \eqref{eq:approx} to the exact expression $\Delta_\phi^2$ in equation \eqref{eq:exact}. We find
\begin{align}\label{eq:adiabatic_approx_ratio}
    \left|\frac{\tilde \Delta_\phi(k)}{\Delta_\phi(k)}\right|
    =
    \sqrt{\frac{c-1}{c}}
    \left|\frac{H^{(1)}_{\nu_\infty}(1)}{H^{(1)}_{\nu_c}\left(\frac{c}{c-1}\right)}\right|
    =
    1 - \frac{\alpha}{c} +\mathcal O\left(\frac{1}{c^2}\right),
\end{align}
where $\alpha = 0.10098...\,$. For large $c$, we can combine this with the expression \eqref{eq:epsilon_power_law} for the first slow-roll parameter $\epsilon$ to obtain the relative error due to the adiabatic approximation in this simple example: 
\begin{align}\label{eq:adiabatic_approx_rel_error}
    \text{relative error} \approx \alpha\epsilon.
\end{align}

There are a couple of things to note regarding the results \eqref{eq:adiabatic_approx_ratio} and \eqref{eq:adiabatic_approx_rel_error}. First, notice that \eqref{eq:adiabatic_approx_ratio} is independent of $k$.
In this simple example of a non-bandlimited power spectrum, the relative error incurred through the use of the adiabatic approximation is the same for all modes. Extrapolating this intuition to our covariantly bandlimited power spectra, this suggests that although the use of the adiabatic approximation likely resulted in small errors in the amplitude and phase of our predicted signal, it seems plausible that our most universal prediction, the frequency of the signal, is unaffected by the use of the adiabatic approximation.

\begin{figure}
    \centering
    \includegraphics[width=0.5\textwidth]{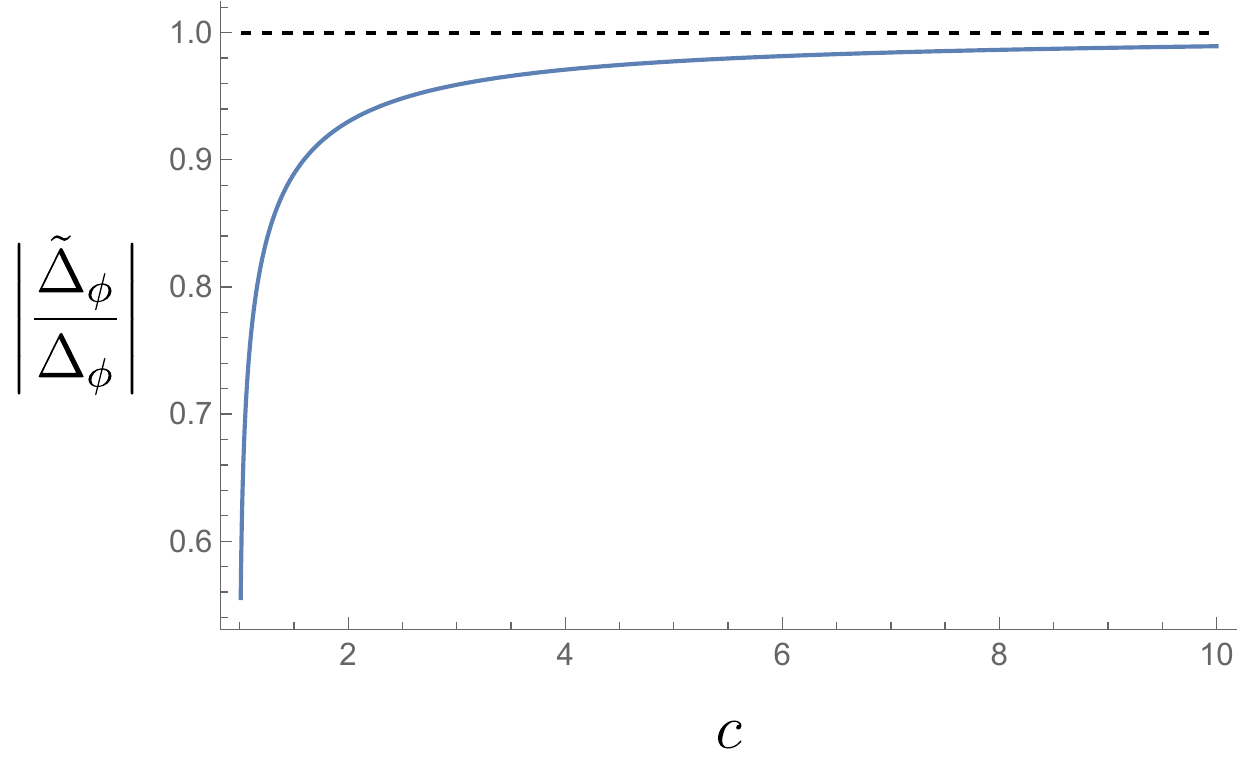}
    \caption{$|\tilde \Delta_\phi(k)/\Delta_\phi(k)|$ versus $c$. $\Delta_\phi(k)$ is the exact value of the scalar power spectrum for a power law spacetime with scale factor $a(t)=A t^c$, while $\Delta_\phi(k)$ is the adiabatic approximation to this quantity. The ratio is independent of $k$ for a power law spacetime.}
    \label{fig:adiabatic_approx}
\end{figure}

Let us now look at the value of $\tilde \Delta_\phi/\Delta_\phi$ as a function of the only variable on which this quantity depends: the power $c$ of the power law expansion. The result is shown in Fig. \ref{fig:adiabatic_approx}. Although we expected that the adiabatic approximation might do well in the $c\rightarrow\infty$ limit, we see that the adiabatic approximation in fact does much better than anticipated. For example, for $c=2$, which is certainly far from a de Sitter expansion, the error due to the adiabatic approximation is only 8\%. Thus, for the slowly rolling spacetimes which we consider in this paper---which \textit{are} very close to de Sitter---we expect that the adiabatic approximation is in fact very accurate. 

As a quantitative estimate of the accuracy of the adiabatic approximation, note that observations put a constraint $\epsilon \ll 0.004$ on the first slow-roll parameter \cite{Planck:2018jri}. From \eqref{eq:epsilon_power_law} we see that to get this small of a value for $\epsilon$ via a power law expansion we require $c>250$. Equation \eqref{eq:adiabatic_approx_ratio}, or the large $c$ approximation \eqref{eq:adiabatic_approx_rel_error}, then give the relative error due to the adiabatic approximation to be only $0.04\%$. Since there seems to be no reason to expect that the adiabatic approximation should do worse when we introduce a covariant cutoff, we estimate that the relative error incurred due to our use of the adiabatic approximation in the covariantly bandlimited case is also likely under one part in one thousand.

\section{Detailed features of the predicted signal}
\label{sec:detailed-features}

After establishing how a covariant natural UV cutoff, $f(\Box)$, is defined in \Eq{eq:cutoff}, we almost exclusively focused on the case of a sharp cutoff, for which $f(\Box) = \theta(\Omega^2 - |\Box|)$.
The class of possible UV cutoffs that one could consider is much larger, however, being parameterized by a functional degree of freedom, $f$.
In principle, $f$ need not even resemble a cutoff; the definition goes through for any non-negative function.
If we insist that $f$ resemble what is normally thought of as a cutoff, however, then we should have $f(\lambda) \approx 1$ for $|\lambda| \ll \Omega^2$, $f(\lambda) \approx 0$ for $|\lambda| \gg \Omega^2$, and a monotonic transition for $|\lambda| \sim \Omega^2$.

Here we investigate how the smoothness and the size of the interval over which $f(\lambda)$ drops from 1 to 0 impacts $\delta \Delta_\phi^2/\Delta_\phi^2$.
While we of course cannot characterize the full functional freedom in $f$, a convenient tool for our investigations is the \emph{smooth step function}:
\begin{equation}
    S_n(x) = \left\{ \begin{array}{ll}
    0 & x \leq 0 \\
    x^{n+1} \sum_{k=0}^n {\binom{n+k}{k}} (1-x)^k & 0 < x < 1 \\
    1 & x \geq 1
    \end{array} \right.
\end{equation}
It has the property that its first $n$ derivatives are continuous at $x = 0$ and $x = 1$, where $n \in \mathbb{Z}$ and $n \geq 0$, and its value increases from 0 to 1 over the interval $0 < x < 1$.
Let us use it to define
\begin{equation} \label{eq:cutoff-smooth}
    f(\lambda) = 1 - S_n\left( \frac{|\lambda| - \Omega^2}{\nu} + 1  \right).
\end{equation}
This profile is a $C^n$ function and it drops from 1 to 0 over the intervals $|\lambda| \in (\Omega^2 - \nu, \Omega^2)$.
The parameter $n$ therefore characterizes the smoothness of the cutoff, and the parameter $\nu$ characterizes the size of the ramp over which the cutoff turns on.

If we replace the sharp profile $f(\lambda) = \theta(\Omega^2 - |\lambda|)$ with $f(\lambda)$ as defined in \Eq{eq:cutoff-smooth}, after percolating through the steps of the calculation, the final consequence is that the integral $I(Q,x)$ changes in the final expression \eqref{eq:final_answer} for $\delta \Delta_\phi^2 / \Delta_\phi^2$.
This is because this integral came from $P_\Omega^\perp \chi_0$; see Eqs.~\eqref{eq:reldiff} and \eqref{eq:dGFah}.
It is convenient to reparameterize the width $\nu$ with a new parameter, $B$, in term of which we replace $I(Q,x)$ with a new integral, $I_n(Q,x,B)$, given by
\begin{equation}
    \begin{aligned}
    I_n(Q,x,B)  &\equiv \int_{0}^\infty \dee q~ S_n\left( \frac{q-Q}{B}+1 \right) q^{-3/2} \cos(q \ln (2q/xe) ) \\
    &= I(Q,x) + \int_{Q-B}^Q \dee q~ S_n\left( \frac{q-Q}{B}+1 \right) q^{-3/2} \cos(q \ln (2q/xe) ).
    \end{aligned}
\end{equation}
In terms of the integration variable $q$, which is related to the eigenvalue $\lambda$ via \Eq{eq:qfunc}, the cutoff smoothly ramps up on an interval $[Q-B,Q]$.
Since $Q \approx \Omega/H = \sigma^{-1}$ when $\Omega \gg H$, it is perhaps most meaningful to think of the ramp-up width $B$ as the number of Planck lengths\footnote{Or more generally cutoff lengths, if the cutoff is at some scale other than the Planck scale.} over which the cutoff turns on if we view the cutoff scale $\Omega$ as being fixed.

\begin{figure}[t]
    \centering
    \includegraphics[width=0.75\textwidth]{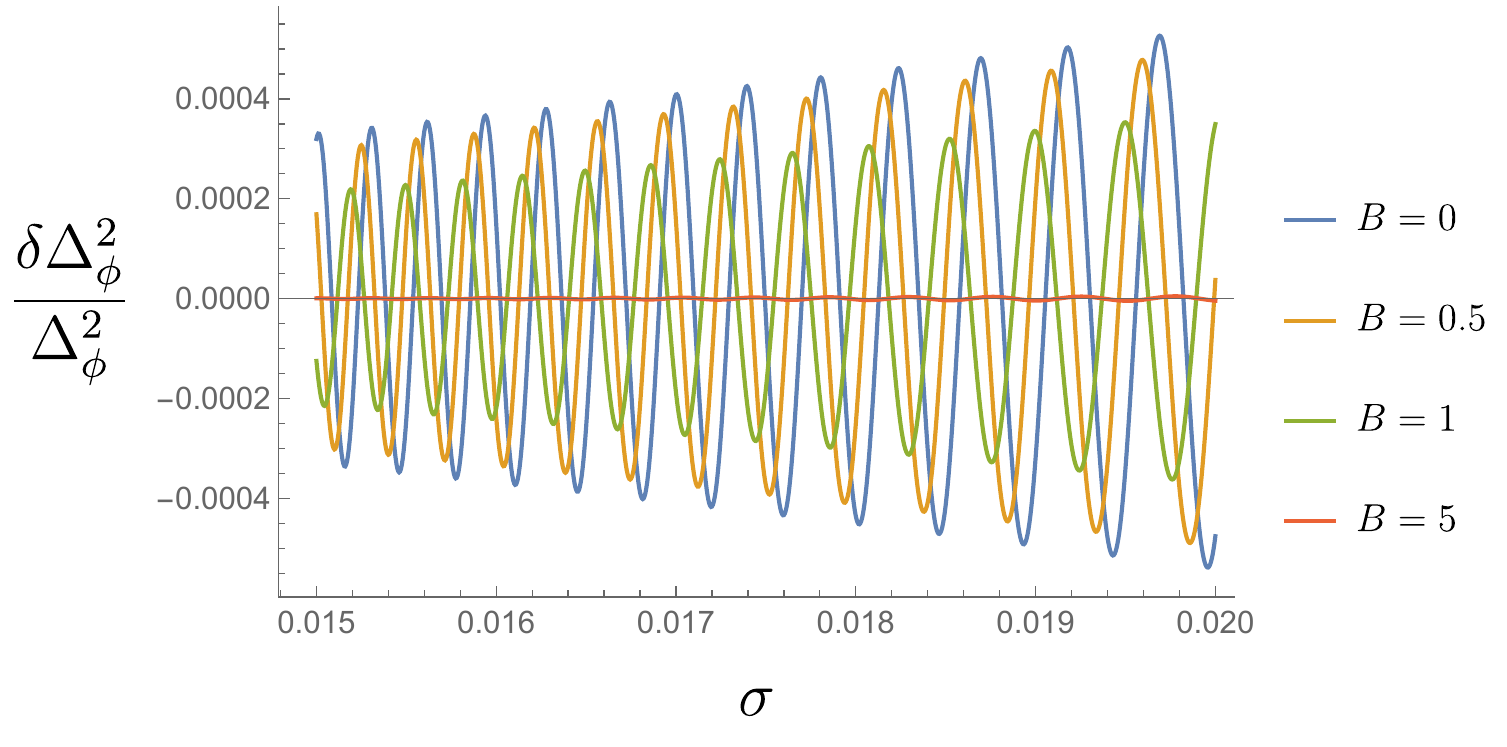}
    \caption{Plots of $\delta \Delta^2_\phi/\Delta^2_\phi$ as a function of $\sigma$ with $x=1$ for various values of the smoothing width, $B$. We have set the smoothness parameter to $n=2$. In this and the following plots, we have that $\sigma \sim 10^{-2}$, which corresponds to a separation between the Hubble and cutoff scales by only two orders of magnitude. Similarly to \Fig{fig:dS-correction}, while this is not very realistic, it makes the features that we wish to highlight easily visible.}
    \label{fig:variable-smoothing-width-fluctuations}
\end{figure}

\begin{figure}[t]
    \centering
    \includegraphics[width=0.6\textwidth]{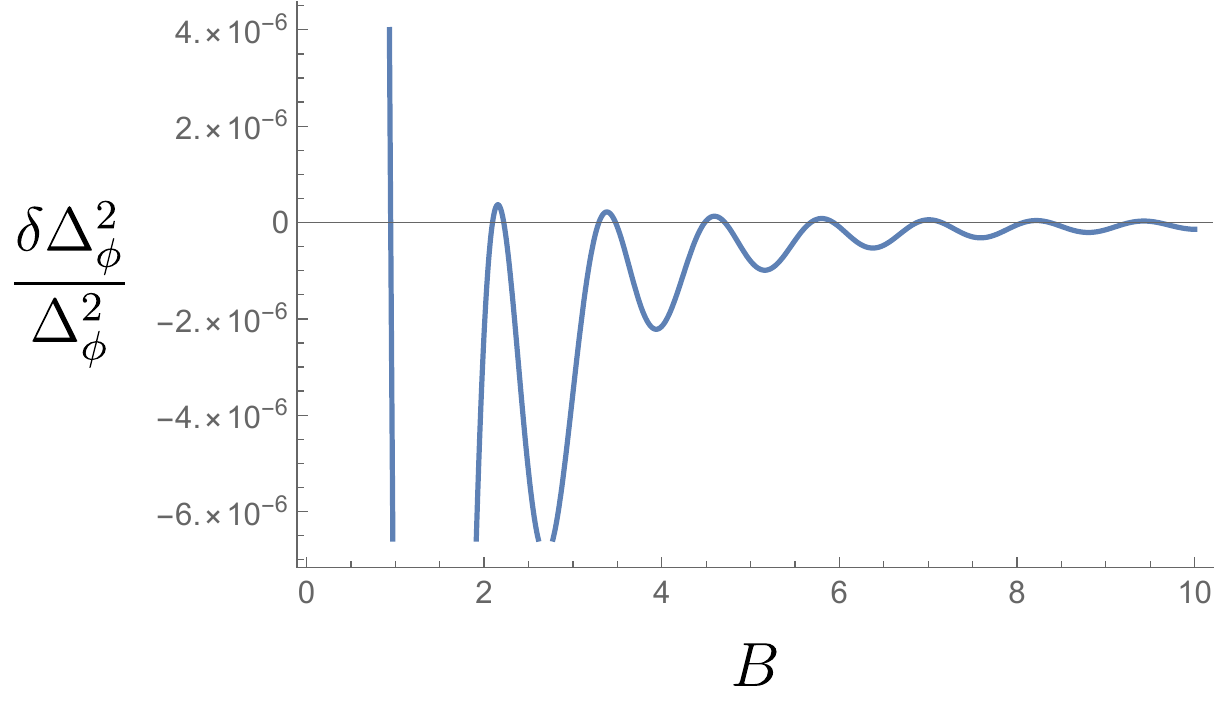}
    \caption{Plot of $\delta \Delta^2_\phi/\Delta^2_\phi$ with $x = 1$ and $\sigma = 10^{-2}$ as a function of the smoothing width parameter, $B$. We have set $n = 2$.}
    \label{fig:smoothing-limit}
\end{figure}

\Fig{fig:variable-smoothing-width-fluctuations} shows what happens to $\delta \Delta_\phi^2 / \Delta_\phi^2$ as we increase the width, $B$.
The general trend is that increasing this width decreases the amplitude of the oscillations and shifts their phase, but it does not seem to change the frequency.
Although increasing $B$ suppresses the signal amplitude, it does not appear to be sending it to zero; see \Fig{fig:smoothing-limit}.
Note that we cannot let $B$ grow too large; otherwise, the approximations made in arriving at the expression \eqref{eq:integrand-approximation} for the integrand of $I_n(Q,x,B)$ break down.

\begin{figure}[t]
    \centering
    \includegraphics[width=0.75\textwidth]{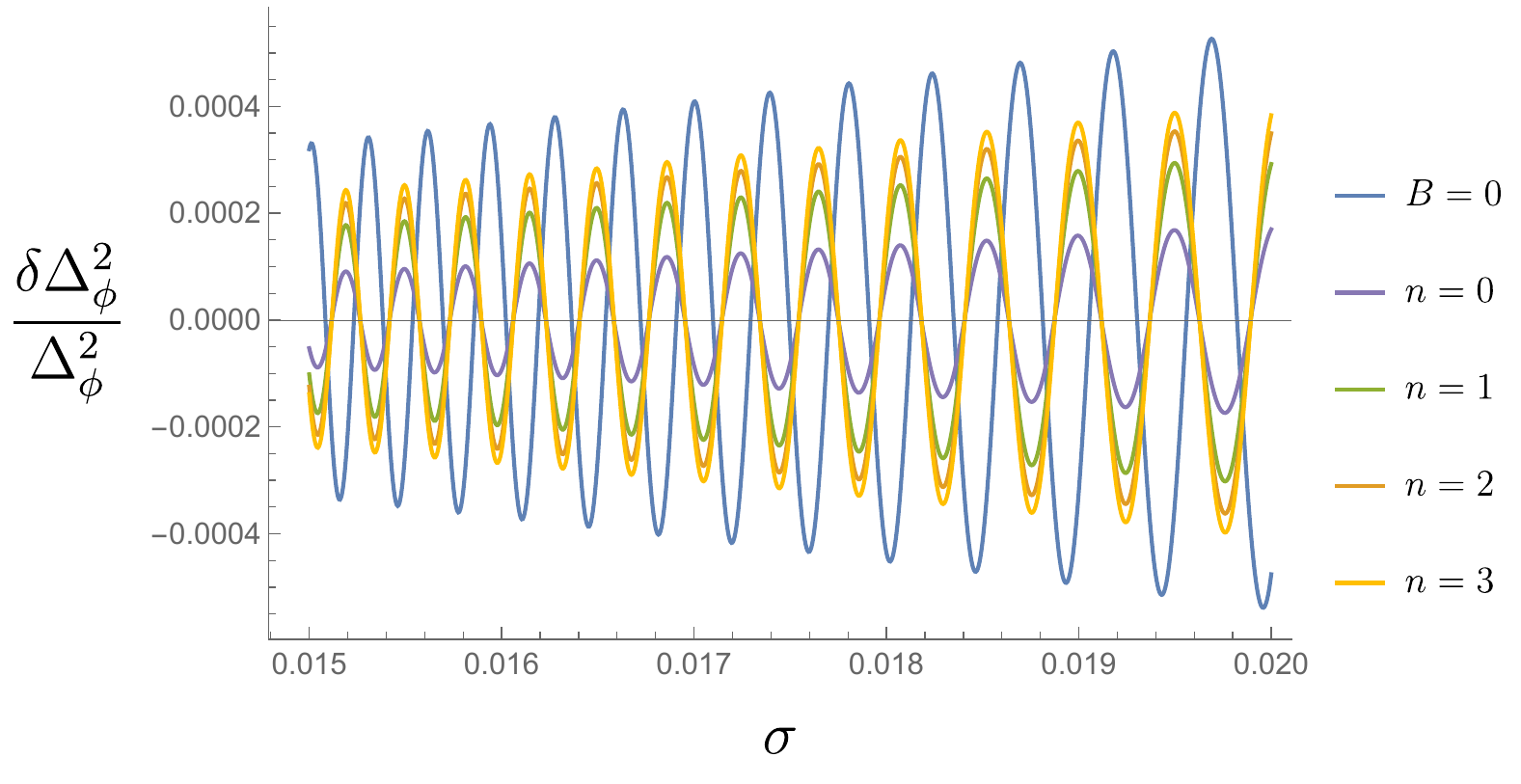}
    \caption{Plots of $\delta \Delta^2_\phi/\Delta^2_\phi$ as a function of $\sigma$ with $x=1$ for various values of the smoothness parameter, $n$. We have set $B=1$ in every case except for that with no smoothing ($B=0$).}
    \label{fig:variable-smoothing-smoothness-fluctuations}
\end{figure}

\Fig{fig:variable-smoothing-smoothness-fluctuations} shows what happens when we adjust the smoothness by changing the parameter $n$.
Again, adjusting the smoothness changes the phase and amplitude, but not the frequency of oscillations.
It furthermore appears that a smooth ramp-up suppresses the amplitude less as it is made smoother.

Of course, precise details of how $\delta \Delta_\phi^2 / \Delta_\phi^2$ changes depends on the choice of smoothing function; however, we expect that the basic qualitative take-home lessons discussed here remain applicable in general.

\clearpage

\bibliographystyle{utphys-modified}
\bibliography{refs.bib}

\end{document}